\documentclass[useAMS,usenatbib]{mn2e}

\usepackage{times}
\usepackage{amsmath}
\usepackage{graphicx}
\usepackage{color}

\newcommand{\Gauss}[1]{\mathcal{N}(#1)}

\newcommand{\ben}{\begin{enumerate}}
\newcommand{\een}{\end{enumerate}}
\newcommand{\bi}{\begin{itemize}}
\newcommand{\ei}{\end{itemize}}
\newcommand{\be}{\begin{equation}}
\newcommand{\ee}{\end{equation}}
\newcommand{\bea}{\begin{eqnarray}}
\newcommand{\eea}{\end{eqnarray}}
\newcommand{\ba}{\begin{array}}
\newcommand{\ea}{\end{array}}
\newcommand{\bc}{\begin{center}}
\newcommand{\ec}{\end{center}}
\newcommand{\bt}{\begin{tabular}}
\newcommand{\et}{\end{tabular}}
\newcommand{\bfig}{\begin{figure}[htb]}
\newcommand{\efig}{\end{figure}}

\newcommand{\bmg}[1]{\mbox{\boldmath $#1$}}
\newcommand{\bm}[1]{ \mathbf{#1} }

\newcommand{\defn}{\stackrel{\mbox{\tiny def}}{=}}

\newcommand{\tp}{^{\sf T}}

\title[Astrophysically robust systematics removal]{Astrophysically
  robust systematics removal using variational inference: application
  to the first month of Kepler data} 

\author[S. Roberts, A. McQuillan, S. Reece \& S. Aigrain]
{S. Roberts$^{1}$\thanks{E-mail: sjrob@robots.ox.ac.uk},
  A. McQuillan$^{2}$, S. Reece$^{1}$ \& S. Aigrain$^{2}$\\
  $^{1}$Department of Engineering Science, University of Oxford,
  Parks Road, Oxford, OX1 3PJ, UK\\
  $^{2}$Department of Physics, University of Oxford, Keble Road, 
  Oxford, OX1 3RH, UK}

\begin{document}

\date{Accepted \ldots Received \ldots; in original form \ldots}

\pagerange{\pageref{firstpage}--\pageref{lastpage}} \pubyear{\ldots}

\maketitle

\label{firstpage}

\begin{abstract}
  Space-based transit search missions such as Kepler are collecting
  large numbers of stellar light curves of unprecedented photometric
  precision and time coverage. However, before this scientific
  goldmine can be exploited fully, the data must be cleaned of
  instrumental artefacts.  We present a new method to correct
  common-mode systematics in large ensembles of very high precision
  light curves. It is based on a Bayesian linear basis model and uses
  shrinkage priors for robustness, variational inference for speed,
  and a de-noising step based on empirical mode decomposition to
  prevent the introduction of spurious noise into the corrected light
  curves. After demonstrating the performance of our method on a
  synthetic dataset, we apply it to the
  first month of Kepler data. We compare the results, which are
  publicly available, to the output of the Kepler pipeline's
  pre-search data conditioning, and show that the two generally give
  similar results, but the light curves corrected using our approach have lower
scatter, on average, on both long and short timescales. We finish by discussing
some limitations of our method and outlining some avenues for further
development.  The trend-corrected data produced by our approach are publicly
available.
\end{abstract}


\section{Introduction}

Surveys for planetary transits monitor tens or hundreds of thousands
of stars at high precision, with rapid time sampling, and over long
periods. The resulting datasets can be used not only to discover and
characterise transiting exoplanets, but also to study a wide range of
phenomena of stellar origin, such as pulsations, rotation, and
binarity.  However, transit survey data are typically affected by a
wide range of artefacts, many of which affect all the target stars
simultaneously, although to varying degrees.

\begin{figure}
\includegraphics[width=\linewidth]{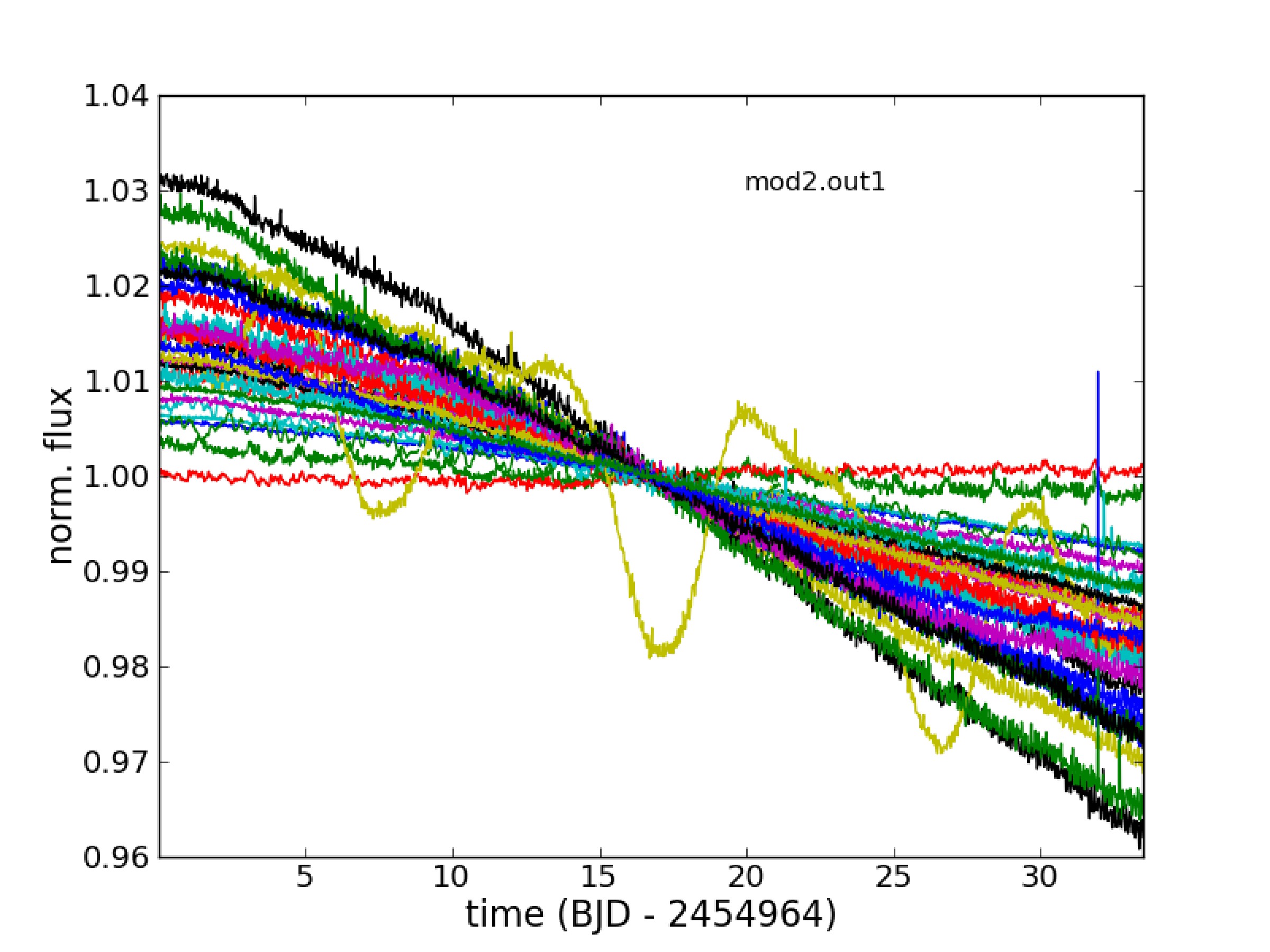}
\caption{Example raw light curves from Kepler quarter 1 data. The 30
  light curves shown were selected at random among the 1290 objects
  observed in the first output channel on module 2 of the Kepler
  detector. Each light curve was normalised by dividing it by its median.}
\label{fig:eg_curves}
\end{figure}

The Kepler space mission, launched on March 6, 2009
\citep{Borucki+:10}, exemplifies this opportunity and the associated
challenges. The unprecedented precision and baseline of the
$\sim$150\,000 Kepler light curves, most of which are now in the
public domain, opens up the possibility of studying the
micro-variability of stars other than the Sun for the first time,
probing amplitudes ranging from a few parts per thousand to a few
parts per million, and timescales of weeks to years. However, there
are also artefacts in the raw data, particularly long term trends,
which are prominent in the light curves of all but the highest
variability stars.  This is visible in Figure~\ref{fig:eg_curves},
which shows a set of example light curves from the first month of
Kepler observations, known as Quarter 1 (hereafter Q1; see
Section~\ref{sec:results_Q1} for more details).

These common-mode instrumental effects are usually known as
\emph{systematics}. When the cause of the systematics is understood well
enough, it is possible to formulate explicit models for their
removal. In many other situations, however, we have little knowledge
regarding the nature or number of systematic trends in the data. In
this paper we present the development of a new methodology, which we
refer to as \emph{Astrophysically Robust Correction} (ARC), which
addresses the issue of discovering and removing underlying systematics
in astronomical data. We apply the method to data from the first
quarter (Q1) of Kepler data, and show that it efficiently removes the
systematic artefacts in that dataset. \cite{McQuillan+:12} used the
resulting 'cleaned' data to perform a preliminary statistical study of
the variability of the Kepler target stars on month timescales.

\subsection{Existing methods}
\label{sec:existing}

The problem of trend identification and removal in ensembles of
stellar light curves was first addressed in the context of
ground-based transit surveys, such as OGLE \citep{Udalski+:02}, HatNET
\citep{Bakos+:02}, or SuperWASP \citep{Pollacco+:06}. The sources of
systematics in these surveys are atmospheric (e.g.\ differential
airmass variations across the field-of-view, seeing and atmospheric
transparency variations) as well as instrumental (e.g.\ pointing
jitter combined with inter- and intra-pixel sensitivity
variations). The two methods, which are most widely used to identify
and remove systematics in ground-based transit surveys are known as
SysRem \citep{Tamuz+:05} and TFA (Trend Filtering Algorithm,
\citealt{Kovacs+:05}).  SysRem attempts to explain each light curve as
a linear superposition of all the other light curves, and adjusts the
coefficients of this linear basis model so as to minimise the total
squared variance of all the light curves. The resulting algorithm is
very similar to Principal Component Analysis (PCA, see e.g.\
\citealt{PracStatAst}), although it allows for individual weighting of
each observation of each star. Because of this, the trends are not
strictly orthogonal, and must be identified and removed
iteratively. The number of times this is done must then represent a
trade-off between removing systematics and preserving real
variability. TFA proceeds in a similar fashion, but the trends are
identified in a limited subset of the light curves, which is selected
by the user, so as to contain a representative sample of the
systematics, but minimal intrinsic variability.  

The images from which the photometry is extracted, or ancillary
sensors, can also be used to measure meteorological and instrumental
parameters, which are thought to affect the photometry, for example
seeing, airmass, detector temperature, etc\ldots From these, one may
construct a linear basis with which to model the systematics, a
procedure known as external parameter decorrelation, (EPD,
\citealt{Bakos+:07}). EPD is usually applied as a preliminary step
prior to running a search for additional trends using (e.g.) SysRem or
TFA. Although the EPD method as described in
  \citealt{Bakos+:07} is designed for large ensembles of light curves,
  the same principle, namely fitting a given light curve with a linear
  combination of external parameters, can also be applied in
  situations where a single star is being monitored, for example
  high-precision observations of planetary transits or eclipses.

All three of the methods outlined above make a number of important
assumptions:
\begin{itemize}
\item that the relationship between the trends and the light curves is linear;
\item that the data used to identify the trends (light curves or
  external parameter data) is sufficient to describe them completely;
\item that the majority of the stars are not significantly variable,
  so that systematics dominate the global sum of the squared residuals
  which is used as a figure of merit.
\end{itemize}
The third assumption becomes problematic in the case of space-based
transit surveys such as Kepler and CoRoT \citep{CoRoT}, which
routinely reach sub-millimagnitude precision. As these missions are
revealing, many, if not most, stars are intrinsically variable at that
level. Intrinsic variability can become an important, if not dominant,
contribution to the global figure of merit one is trying to optimize,
exacerbating the problem of preserving this variability whilst
removing the systematics. As a result, in PCA-like
  approaches, the first few principal components can easily be
  dominated by a small number of highly variable stars, which happen
  to account for a large fraction of the overall variance of the
  data. This problem is exacerbated by the fact that these methods are
  couched in the framework of maximum likelihood, and therefore are
  not robust against overfitting. \citet{Petigura+Marcy:12} partially
  address this problem, by identifying and removing from the
  training set any light curves, whose basis coefficients are deemed
  anomalous. In this paper we go further, introducing a criterion
  which explicitly expresses our desiderata that the trends should be
  constituted of many small contributions from many light curves, and
  using a fully Bayesian model to avoid over-fitting (see
  Section~\ref{sec:ARCintro}).

  The pipeline used to process Kepler data before releasing it to the
  public incorporates a `pre-search data conditioning' (PDC) stage.
  The PDC is designed to remove instrumental systematics so as to
  optimize the efficiency of the transit detection process. Early
  implementations of the PDC \citep{Jenkins+:10} followed a procedure
  similar to EPD, but attempted to identify and remove signals of
  stellar origin in the frequency domain before applying the
  correction. This preserved astrophysical variability in some, but
  not all cases. Additionally, in some cases the PDC was found to
  introduce high-frequency noise into the light curves. This effect
  has been noted in a previous publication \citep{murphy:12} in which
  the noise properties of short (1-min) and long (30-min) cadence data
  were compared for the same targets. It is due to the fact that the
  estimate of each systematic trend is necessarily noisy.  These
  issues made the PDC-processed light curves unsuitable for
  variability studies, and motivated the development of the
  alternative method, which is the subject of this paper. In the mean
  time, Kepler pipeline has also evolved. In more recent
  implementations of the PDC, known as PDC-MAP (maximum a-posteriori,
  \citealt{PDC-MAP1,PDC-MAP2}), a subset of highly correlated and
  quiet stars is used to generate a cotrending basis vector set, as
  well as a set of priors over the coefficients of each basis vector,
  which is then used in the correction of all the light curves.  This
  leads to a significant improvement in robustness, but we shall defer
  a detailed comparison to Section~\ref{sec:results_Q1}.

\subsection{Introducing the ARC}
\label{sec:ARCintro}

Our starting point in the development of a robust detection and
correction methodology for the Kepler data was not fundamentally
different to those of previously published methods: like virtually
all of them, we use a linear basis model and identify
the trends from the light curves themselves, as is done in SysRem and
TFA. However, our approach differs in the following important ways:
\begin{itemize}
\item we perform the linear basis regression using the \emph{Variational
  Bayes} (VB) method of approximate Bayesian inference to ensure that
  the procedure is robust to uncertainties in the data, while
  maintaining computational tractability and scalability to large
  datasets;
\item we apply \emph{automatic relevance
      determination(ARD)} priors to the coefficients of the basis
    vectors. These priors have zero mean and their individual widths
    are adjusted iteratively to maximise the model evidence. This
    ensures that trends are identified and removed only if there is
    significant evidence for them in the data;
\item we formulate an explicit criterion, based on a measure of the
  \emph{entropy} of each trend, to reflect our belief that the trends are
  systematic i.e.\ that they are present, at some level, in the
  majority of light curves, and that the contribution of any single
  light curve to any given trend should be small;
\item we incorporate a de-noising step after the identification of
  each trend, which uses the \emph{empirical mode decomposition} (EMD)
  algorithm, chosen to avoid any intrinsic bias towards trends on
  particular timescales.
\end{itemize}
The entropy criterion is similar in spirit to
  the modified PCA method of
  \citet{Petigura+Marcy:12}, but it actively selects trends which
  satisfy our desiderata, rather than merely rejecting problematic
  light curves.
\smallskip

This paper is structured as follows. Section \ref{sec:method} gives an
overview of the ARC, focussing on aspects which are new and/or with
which our intended readership may be unfamiliar, namely our trend
selection and stopping criteria and the EMD
algorithm. Section~\ref{sec:results} shows some illustrative results,
starting with a walk-through on a synthetic example dataset, and then
moving on to the Kepler Q1 data. In Section~\ref{sec:disc}, we compare
the performance of the PDC-MAP and the ARC.
Finally, Section~\ref{sec:conclusions}
summarises the advantages and limitations of the ARC and discusses open
questions on which future work will focus. Detailed descriptions of the
Bayesian linear basis regression model and of the VB inference method
are given for reference in Appendices~\ref{sec:vblbm} and
\ref{sec:VB}.

\section{Method}
\label{sec:method}

The main challenge, when performing systematics removal
  with a linear basis model, is to identify the appropriate basis,
  i.e.\ the systematic trends. We adopt an iterative approach to this
  problem, isolating the most dominant systematic trend first,
  removing it and then repeating the process. At each iteration, the
  dominant systematic is identified by constructing a set of
  `candidate' trends from the light curves themselves, ranking them
  according to the entropy criterion, selecting the top few
  candidates, combining them into a single time-series using PCA, and
  denoising the resulting trend. We now proceed to describe each of
  these steps in detail.

First, we introduce a few notation conventions.  Our data consists of
$N$ observations each of the relative flux of $M$ stars. Throughout this
paper, we use the subscript $n (= 1$, \ldots, $N)$ to denote
observation number and $m (= 1$, \ldots, $M)$ to denote the star
number. We denote the $n^{\rm th}$ observation of star $m$ as
$d_{nm}$, and the corresponding observation time $t_n$. The light
curve of star $m$ forms a column vector $\bm{d}_m \equiv [d_{1m}$,
\ldots, $d_{Nm}]\tp$ (where $\bm{A}\tp$ is the matrix transpose of
$\bm{A}$). 

We consider each of the observed light curves, $\bm{d}_m$, to be composed of
a linear combination of an underlying `true' stellar signal, $\bm{s}_m$,
an unknown number, $K$, of systematic trends, $\bm{u}_k$ ($k=1$,
\ldots, $K$), and an observation noise process, $\bmg{\epsilon}_m$:
\begin{equation} 
\bm{d}_m = \bm{s}_m + \sum_{k=1}^{K} a_{mk} \bm{u}_k + \bmg{\epsilon}_m
\label{eq:light_curve}
\end{equation}
where the unknown factors $a_{mk}$ represent the contribution of
the $k^{\rm th}$ systematic trend to the $m^{\rm th}$ light
curve. 

Defining the ensemble of all other light curves, $\bm{D}_m =
\{\bm{d}_{l \not= m} \}$, we seek to model as much of $\bm{d}_m$ as
possible using $\bm{D}_m$ as a \emph{basis}. This is achieved via a
Bayesian linear model with inference performed using variational Bayes
(see Appendices~\ref{sec:vblbm} and \ref{sec:VB} for details) such
that,
\begin{equation}
\bm{\hat{d}}_m = \sum_{l \not= m} w_{ml} \bm{d}_l
\label{eq:linMod}
\end{equation}
where the $\{ w_{ml} \}$ are a set of weights. We repeat the process
for each light curve in turn, resulting in a set of putative
explanatory vectors $\{\bm{\hat{d}}_m\}$. In the remainder of this
section, we refer to the $\bm{\hat{d}}_m$ as candidate trends, to
distinguish them from the `adopted' trends $\bm{u}_k$.

\subsection{Trend entropy}
\label{sec:entropy}

Our primary hypothesis is that systematic trends are present in the
majority of the light curves. Thus, if $\bm{\hat{d}}_m$ represents a
linear combination of `true' systematic trends, it should be composed
of many small contributions from many of the other light curves,
rather than a few dominant contributions from a small number of light
curves. More formally, the distribution of the weights $\bm{w}_m =
\{w_{ml}\}$ should have a high Shannon entropy \citep{shannon_ent}:
\begin{equation}
\mathcal{H}\left(\bm{w}_m\right) = -\sum_{l \not=m} p_{ml} \log_2 p_{ml},
\label{eq:entropy}
\end{equation}
where we have defined
\begin{equation}
p_{ml} = \frac{w^2_{ml}}{\sum_{l' \not= m} w^2_{ml'}}
\end{equation}
by analogy with a normalised probability distribution.

\begin{figure}
\includegraphics[width=\linewidth]{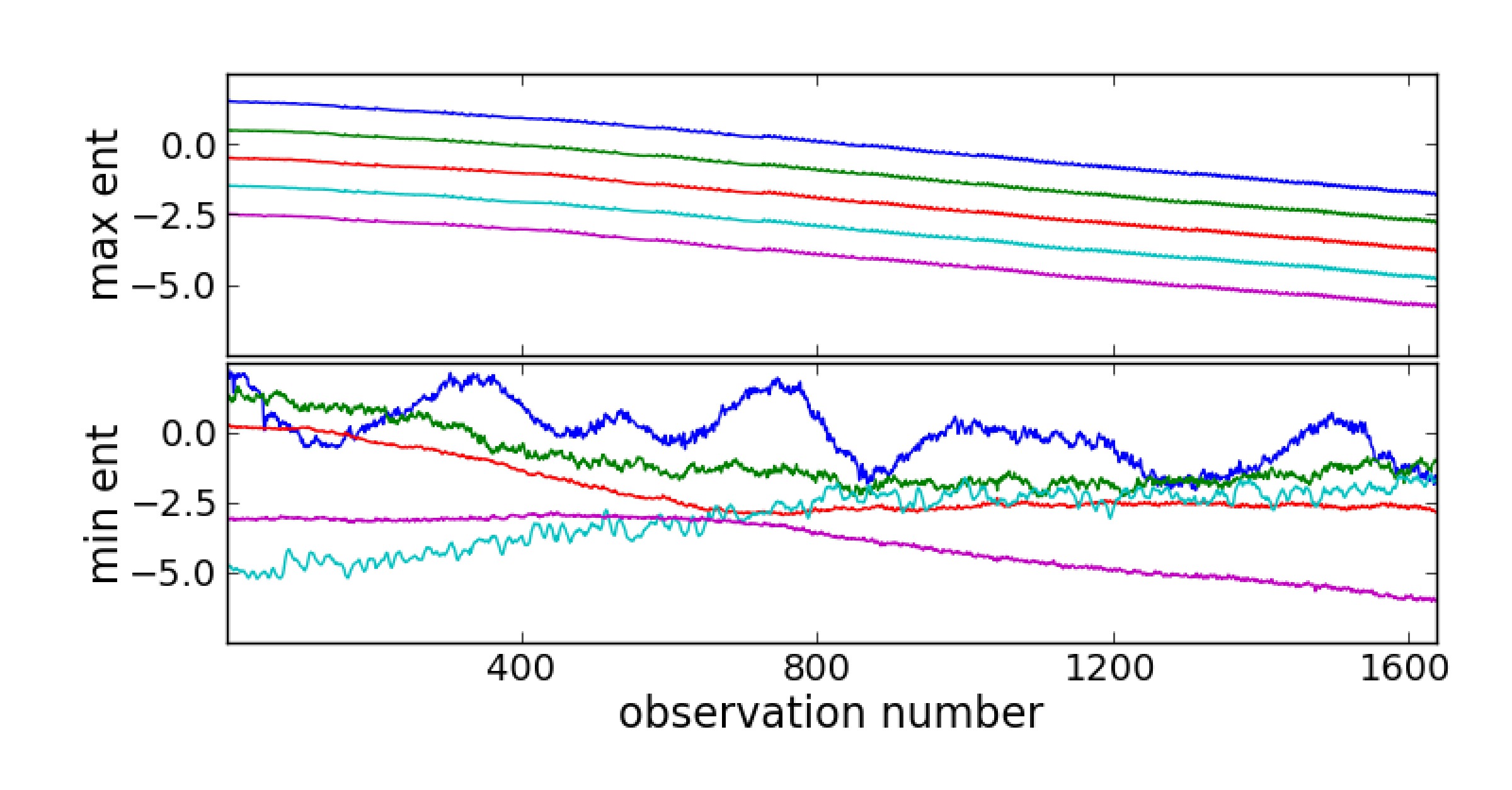}
\caption{Maximum (top) and minimum (bottom) entropy candidate trends
  for the Kepler Q1 data (using all of the Q1 light curves). The
  trends have been normalised to have zero mean and unit
  variance. Vertical offsets have been added and only five curves
  shown in each category to improve figure clarity. }
\label{fig:maxent_basis1}
\end{figure}

We therefore rank each candidate trend $\bm{\hat{d}}_m$ according to
the entropy $\mathcal{H}\left(\bm{w}_m\right)$ of the associated set
of weights. The $\bm{\hat{d}}_m$ with the highest entropy are expected
to be mutually similar, since they all represent a \emph{systematic}
trend. On the other hand, those with the lowest entropy correspond to
cases where a particular light curve was found to be very similar to
one or two others, but not to the rest. This is illustrated in the
case of the Kepler Q1 data by Figure \ref{fig:maxent_basis1}. We form
a reduced basis $\bm{T}$ by selecting the few candidate trends with
the highest entropy. We normally use ten, but as the maximum entropy
trends are mutually similar, the exact number (within reason) is not
important. 

\subsection{Principal component analysis and spectral radius}

If, as expected, the trends in the reduced basis
  $\bm{T}$ are representative of a single, dominant systematic, then
  they should be mutually similar, as illustrated in
  Figure~\ref{fig:maxent_basis1}. We can thus use PCA to extract a
  single trend from the reduced basis (the first principal component), and evaluate just how dominant
  this trend is. The first principal component of $\bm{T}$
  contains most of the information, and explains most of the overall
  variance of $\bm{T}$.

We start by decomposing $\bm{T}$ into a pair of orthonormal matrices
of singular components, $\bm{U}$, and $\bm{V}$, and a diagonal matrix
$\bm{S}$ of singular values, $\sigma_i$. The latter are the square
roots of $\lambda_i$, the eigenvalues of $\bm{TT}\tp$.
\begin{equation}
  \bm{T} = \bm{USV}\tp.
  \label{eq:PCA}
\end{equation}
$\bm{U}$ is also the matrix of eigenvectors of $\bm{TT}\tp$:
\begin{equation}
\bm{TT}\tp = \bm{U}\bm{S}\bm{S}\tp\bm{U}\tp = \bm{U}\bmg{\Lambda}\bm{U}\tp,
\label{eq:eig}
\end{equation}
in which $\bmg{\Lambda}$ is a diagonal matrix of eigenvalues,
$\{\lambda_i\}$. The \emph{spectral radius} of each eigenvector,
defined as
\begin{equation}
\rho_i = \frac{\lambda_i}{\sum_j \lambda_j},
\label{eq:specR}
\end{equation}
is then a measure of the fraction of the overall variance of $\bm{T}$
explained by that eigenvector. In this paper we consider, in
particular, the spectral radius,$\rho_1$, of the first eigenvector.

\begin{figure}
  \includegraphics[width=\linewidth]{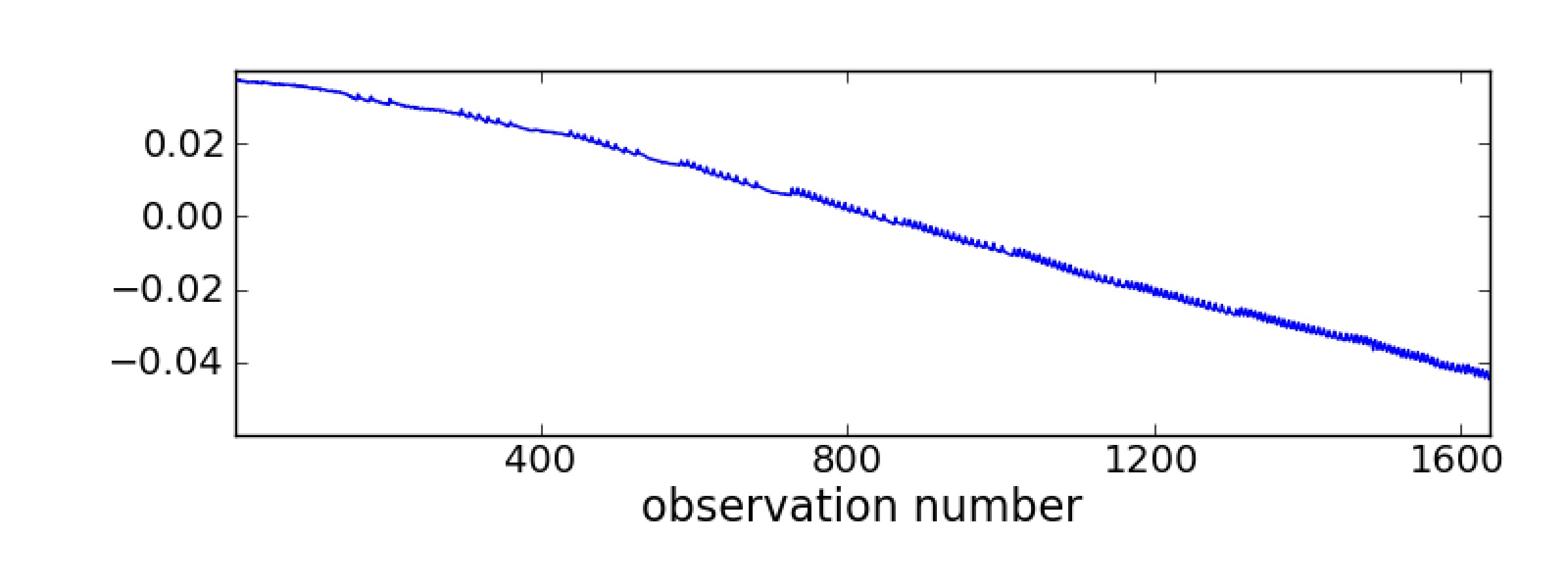}
  \caption{First principal component for the set of five
    maximum-entropy candidate trends shown in the top panel of
    Figure~\ref{fig:maxent_basis1}. }
\label{fig:u_eg}
\end{figure}

As discussed above, if the candidate trends which make up $\bm{T}$ are
truly systematic, they should also be \emph{self-similar}, and the
spectral radius of the first principal component should be large
(close to unity). If so, the first principal component contains
pertinent information regarding the systematic trends in the data. For
example, Fig.~\ref{fig:u_eg} shows the first principal component of
the matrix $\bm{T}$ identified from all the Kepler Q1 light
curves. Its spectral radius is $>0.99$, so it is highly representative
of the $\bm{T}$, and it would be reasonable to adopt it as one of the
$\bm{u}_k$. However, as it is a linear combination of a finite number
of noisy light curves, it may still contain significant amounts of
noise. Using it directly may therefore introduce noise into the light
curves, particularly those of bright stars. Thus, we introduce a final
de-noising step in our trend discovery process.

\subsection{Empirical Mode Decomposition (EMD)}
\label{sec:emd}

\emph{Empirical Mode Decomposition} (EMD, \citealt{Huang}) decomposes
a time-series into a series of \emph{intrinsic modes}, each of which
admits a well-behaved Hilbert transform, enabling the computation of an
energy-frequency-time distribution, or Hilbert spectrum. We chose to
use EMD for de-noising of the systematic trends, rather than any other
method, because it does not focus on a particular frequency range, but
it also does not rely on harmonicity or linearity.

\begin{figure}
  \includegraphics[width=\linewidth]{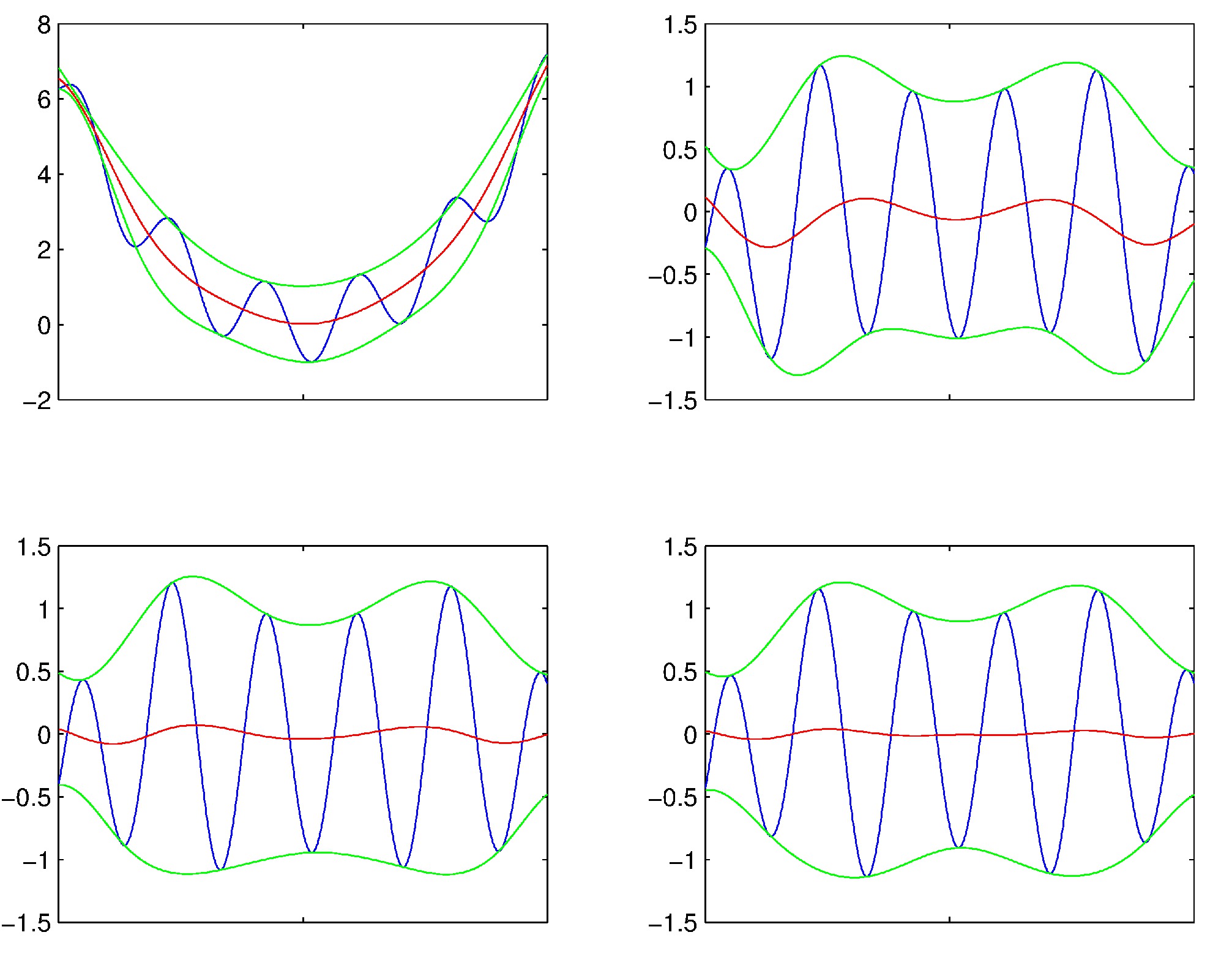}
  \caption{The sifting process of EMD shown over four iterations for
    an example synthetic dataset (top left, top right, bottom left,
    then bottom right). The original time series (blue, top left) is
    successively deflated via a baseline (red curve) formed as the
    mid-point of the envelope functions (green curves). This process
    repeats until changes in the baseline curve are below threshold.}
 \label{fig:emd_eg}
\end{figure}
 
The core computation of EMD lies in the successive detection of maxima
and minima in the time-series. These turning points are thence used to define upper and
lower envelopes to the signal by fitting a spline curve which
passes through the turning-point locations. The mid-point of these
envelopes is then used as a baseline which is removed from the
signal. This procedure is iterated until the baseline function is flat
(to within a threshold). The resultant waveform at this point forms an
intrinsic mode, which is then removed from the original time
series. The entire process is then repeated, successively extracting
intrinsic modes until the residual has no turning points.

\begin{figure}
  \centering
  \includegraphics[width=0.5\linewidth]{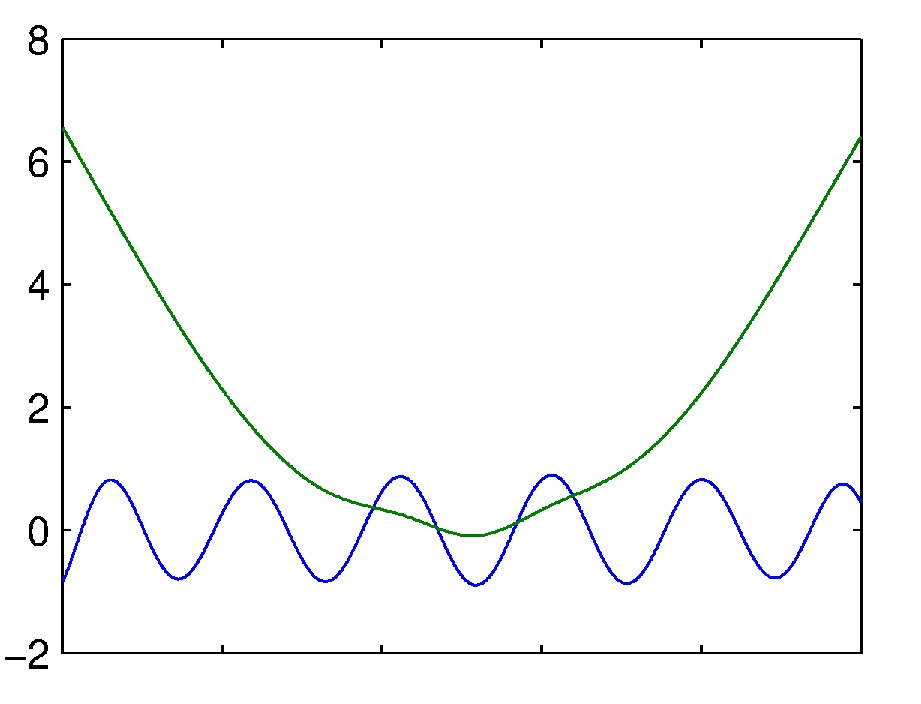}
  \caption{The two resultant intrinsic modes for the simple sine and
    quadratic example shown in Fig.~\protect\ref{fig:emd_eg}.}
  \label{fig:emd_eg_res}
\end{figure}

Figure \ref{fig:emd_eg} shows this process, which is referred to as
\emph{sifting} in the EMD literature, applied to a simple synthetic
time-series. The example signal consists of a sine curve superimposed
on a quadratic, as shown in the blue trace in the top left
sub-figure. The green curves are the envelope functions at each
iteration and the red trace is the mid-point baseline, which is
subtracted from the time series at each iteration. We see that, even
after a small number of iterations, the sine component is
highlighted and this indeed forms the first intrinsic mode of the
data. We continue this process until relative changes in the baseline
function are $<10^{-3}$ (this results in some 20 iterations for the
example shown here).  Figure \ref{fig:emd_eg_res} shows both
resultant intrinsic mode functions extracted from this data after
iterating the sifting process until relative changes are below
the threshold.

\subsection{Iterative trend identification and removal}
\label{sec:id_removal}

Starting from the original set of light curves, the trend
identification and removal proceeds as follows:
\begin{enumerate}
\item identify a set of candidate trends $\{\bm{\hat{d}}_m\}$, each
  associated with entropy $\mathcal{H}(\bm{w}_m)$;
\item construct a candidate basis $\bm{T}$ from the 10 candidate
  trends with the highest entropy;
\item apply PCA to $\bm{T}$ and measure the spectral radius $\rho_1$
  of its first principal component;
\item if $\rho_1 < \rho_{\rm min}$, stop.
\item otherwise, perform EMD on the first principal component, and
  adopt the intrinsic mode with the largest variance as the next
  systematic trend $u_k$;
\item remove all the trends identified so far to produce a set of
  partially corrected light curves
  \begin{equation}
    \bm{\tilde{d}}_m = \bm{d}_m - \sum_{k=1}^{K}
    a_{mk} \bm{u}_k,
  \end{equation}
  where the $a_{mk}$ are once again found using VB with ARC priors.
\item return to step (i), with the $\{\bm{\tilde{d}}_m\}$ as inputs;
\end{enumerate}
The process continues until $\rho_1 < \rho_{\rm min}$ at step
(iv). 

The choice of stopping threshold $\rho_{\rm min}$ is a matter of
fine-tuning. Typical values used in this work varied between 0.6 and
0.9, and the appropriate threshold was found by trial and error (i.e.\
by examining small subsets of the corrected light curves obtained
using various values of $\rho_{\rm min}$). All the Kepler results
presented in this paper used $\rho_{\rm min}=0.8$. We note
  that one could also select this threshold based on theoretical
  arguments. For example, recalling that the $\rho_i$'s are normalised
  eigenvalues of the matrix $\bm{TT\tp}$ (Eqs.~\ref{eq:eig} \&
  \ref{eq:specR}), we can make use of the fact that the set of
  eigenvalues of an arbitrary matrix are bounded by the Weyl
  inequalities. For a large matrix, this can be extended to formulate
  probability distributions over the expected values in the presence
  of noise \citep{Everson+Roberts:00}. This could in principle be used
  to select $\rho_{\rm min}$ in a probabilistic manner. However,
  ultimately, the threshold will remain subjective, whether it is
  couched in terms of $\rho$, or a user-defined credibility limit in
  probability space. On balance, we felt that trial and error was the
  best means of ensuring that the choice of $\rho_{\rm min}$ reflects
  the all-important domain knowledge of the user.

The computational cost of the VB linear basis modelling done in step
(i) scales as $M^2N$. It thus becomes prohibitive if $M$ is very
large. When treating a large number of light curves, it is possible to
decouple the trend identification and removal processes, using only a
subset of the light curves for trend identification. This subset might
be selected at random, or adjusted to contain light curves displaying
evidence of systematics, but not strong intrinsic variability. The
trend removal process scales more benignly as $MN$ and so can be
applied easily even to large sets of light curves.

\section{Results}
\label{sec:results}

In this section we apply the ARC method outlined above to two example
datasets: first, a synthetic example for explanatory purposes, and the
Kepler Q1 data, where we compare our results to the output of the
existing PDC-MAP pipeline.

\subsection{Synthetic example}
\label{sec:synth_eg}

In this subsection we present a simple synthetic example, to act as a
`walk-through' guide to the methodology and as a proof of concept.

\begin{figure}
  \centerline{\includegraphics[width=0.9\linewidth]{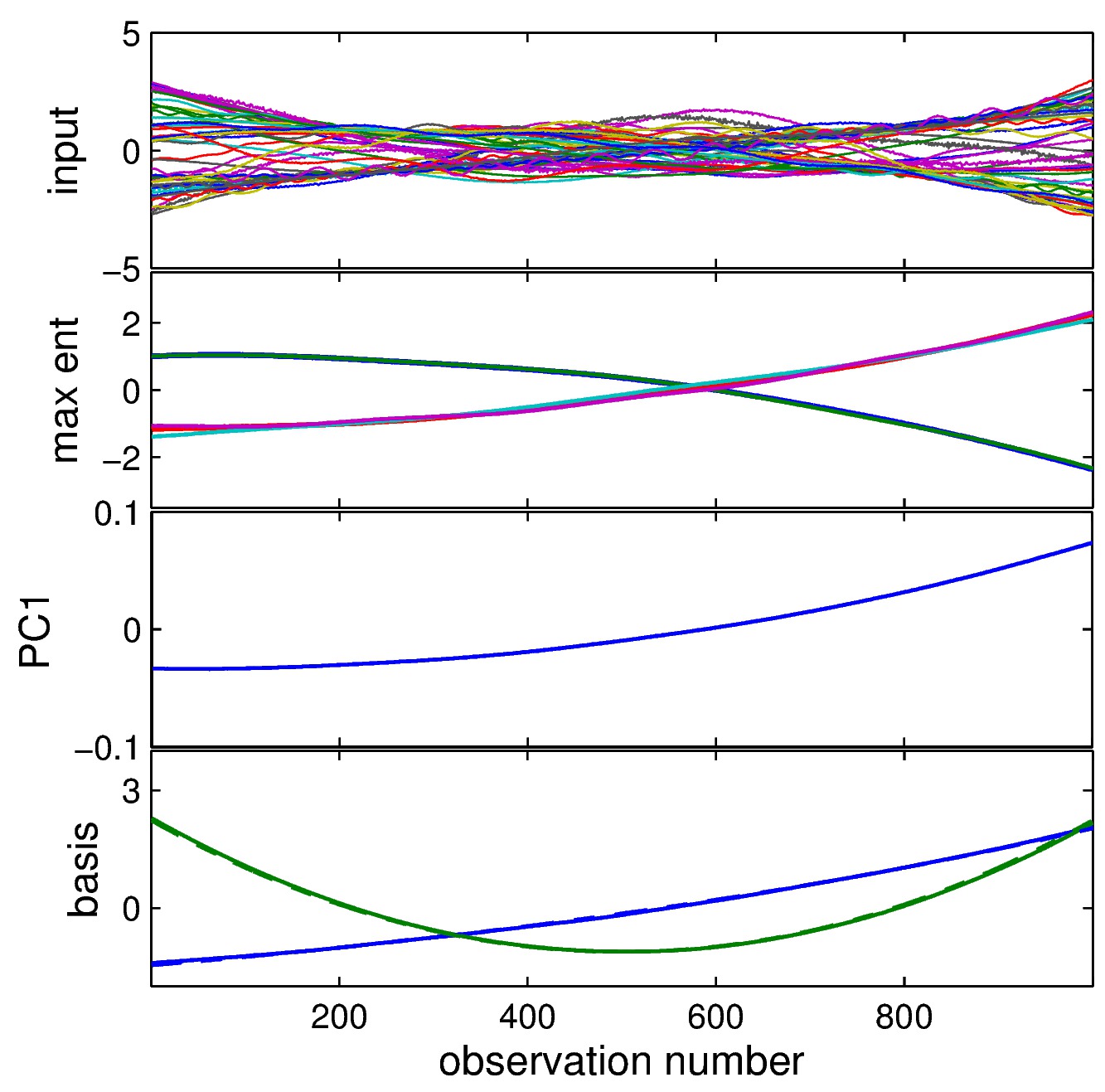}}
  \caption{\protect Application of the ARC to a synthetic dataset. Top panel:
    randomly chosen subset of input light curves. Second panel: the five
    highest entropy candidate trends found during the first
    iteration. Third panel: the first principal component of the
    high-entropy trends, which accounts for $>95$\% of the
    variance. Bottom panel: the two basis trends discovered by the
    algorithm (solid lines) and the corresponding trends injected into
    the data (dashed lines); we note that the solid and dashed lines are
    almost undistinguishable as the recovered trends are very close to the originals.}
 \label{fig:synth_eg}
\end{figure}

We start with a set of 200 synthetic `light curves' formed from
random phase, frequency and amplitude sine
  curves, plus additive Gaussian noise. To these we add random
amounts of two `systematic trends': an exponential decay with long
time constant, and a quadratic term. This constitutes our input
dataset, examples of which may be seen in the top panel of Figure
\ref{fig:synth_eg}. Note that, without loss of generality, we
normalise the curves such that each have unit variance and zero mean.

\begin{figure}
 \centerline{\includegraphics[width=0.9\linewidth]{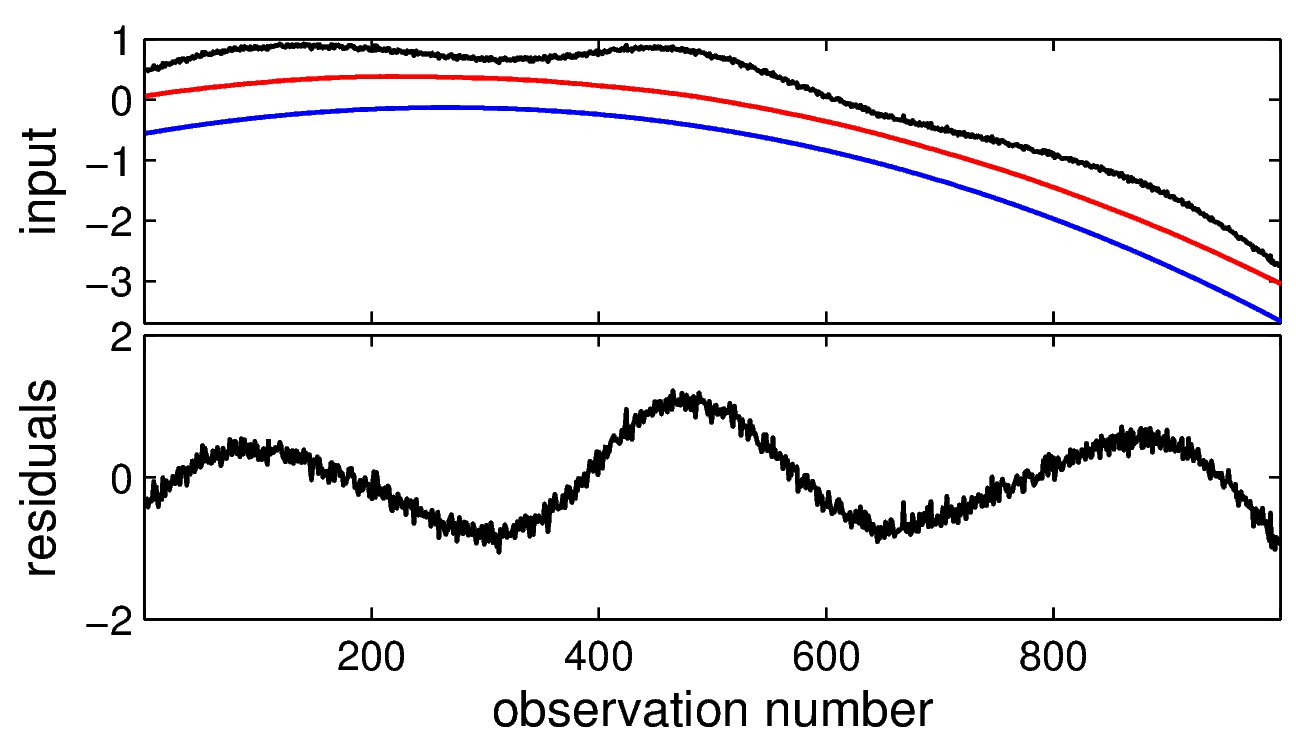}}
 \caption{\protect Result of the ARC
   applied to a synthetic dataset. Top panel: example `light curve',
   with the systematics model (linear combination of the two
   identified trends) shown in red, and the actual trends injected in
   blue (with a vertical offset for clarity). Bottom panel: residuals
   after trend removal using the identified trends. }
 \label{fig:synth_eg_res}
\end{figure}

The steps of the method are illustrated in Figure \ref{fig:synth_eg}.
The trends were identified in a representative subset of 50 `light
curves' chosen at random, and the basis $\bm{T}$ was constructed from
the ten highest-entropy trends at each iteration. Two systematic trends
were identified before the stopping criterion was reached using
$\rho_{\rm min}=0.6$. The resulting correction is illustrated on an
example `light curve' (not one of the subset in which the trends were
identified) in Figure~\ref{fig:synth_eg_res}.

In this synthetic example we may compare the original data, prior to
addition of systematic trend artefacts, with the recovered detrended
traces, as we have the luxury of knowing the `ground truth'. The
resultant linear correlations between the true `light curves' and
those recovered by the algorithm lie between 0.94 and 0.999, with a
median value of 0.98, indicating a very high reconstruction accuracy
(we note that the median correlation between the original curves and
the trend-corrupted versions is only 0.11).

\subsection{Application to Kepler Q1 data}
\label{sec:results_Q1}

We now apply the ARC to the first month of Kepler data, known as
Q1. The ARC-corrected Q1 data were already used by
\citet{McQuillan+:12} to study the statistics of stellar variability,
but here we present a detailed description of the results and a
comparison with the more recently released PDC-MAP data. The PDC-MAP
data we use in this paper were made public on 25 April 2012 as part of
Data Release 14, although the only significant difference between this
and the earlier release of Q1 used by \citet{McQuillan+:12} is the
replacement of the original PDC results with the output of the newer
PDC-MAP pipeline. The data to which the ARC was applied are
essentially the same.

The Q1 data consist of 1639 observations of 156\,097 stars taken over
33.5 days in May--June 2009. The data we consider has been
pre-processed only to convert it from pixel-level data to light
curves, as described in \citet{Jenkins+:10}. The incident light is
collected by 24 square modules, arranged to cover the focal plane of
the telescope \citep{kepInstHandbook}. Each module consists of two
rectangular charge-coupled device (CCD) detectors, each of which has
two readout channels, resulting in a total of 84 channels and 94.6
million pixels.

\begin{figure}
\includegraphics[width=\linewidth]{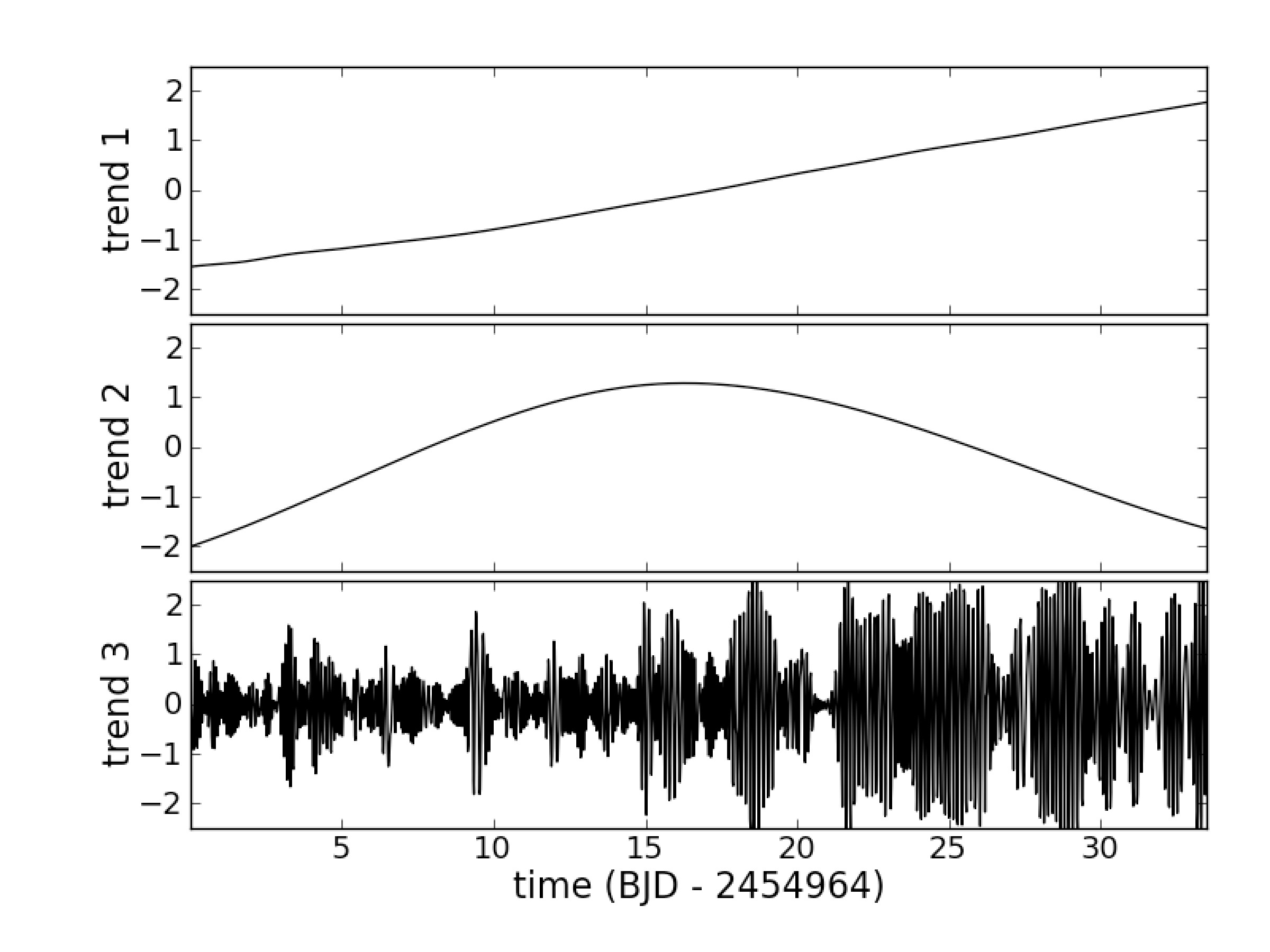}
\caption{Basis trends inferred from the global set of all Kepler Q1
  light curves.}
\label{fig:global_basis}
\end{figure}
We first identified a single set of basis trends for all the light
curves using a randomly selected subset of 200 light curves and using
a threshold $\rho_{\rm min}=0.8$. These are shown in Figure
\ref{fig:global_basis}. The first two have low-complexity, and are
roughly linear and quadratic in nature, respectively. The third shows
more complex dynamics, but is nonetheless present in a very large
number of Q1 light curves. It is caused by the on/off switching of the
reaction wheel
heaters \citep{DataChar}.
We note that, although the first two trends resemble
  low-order polynomials, they are \emph{not} simply linear and
  quadratic components. Furthermore, if we were to remove such
  low-order polynomials from each light curve individually, we would be failing to
  take into account the fact that the systematic trends are
  \emph{global} rather than specific to each light curve.

\begin{figure*}
  \includegraphics[width=\linewidth]{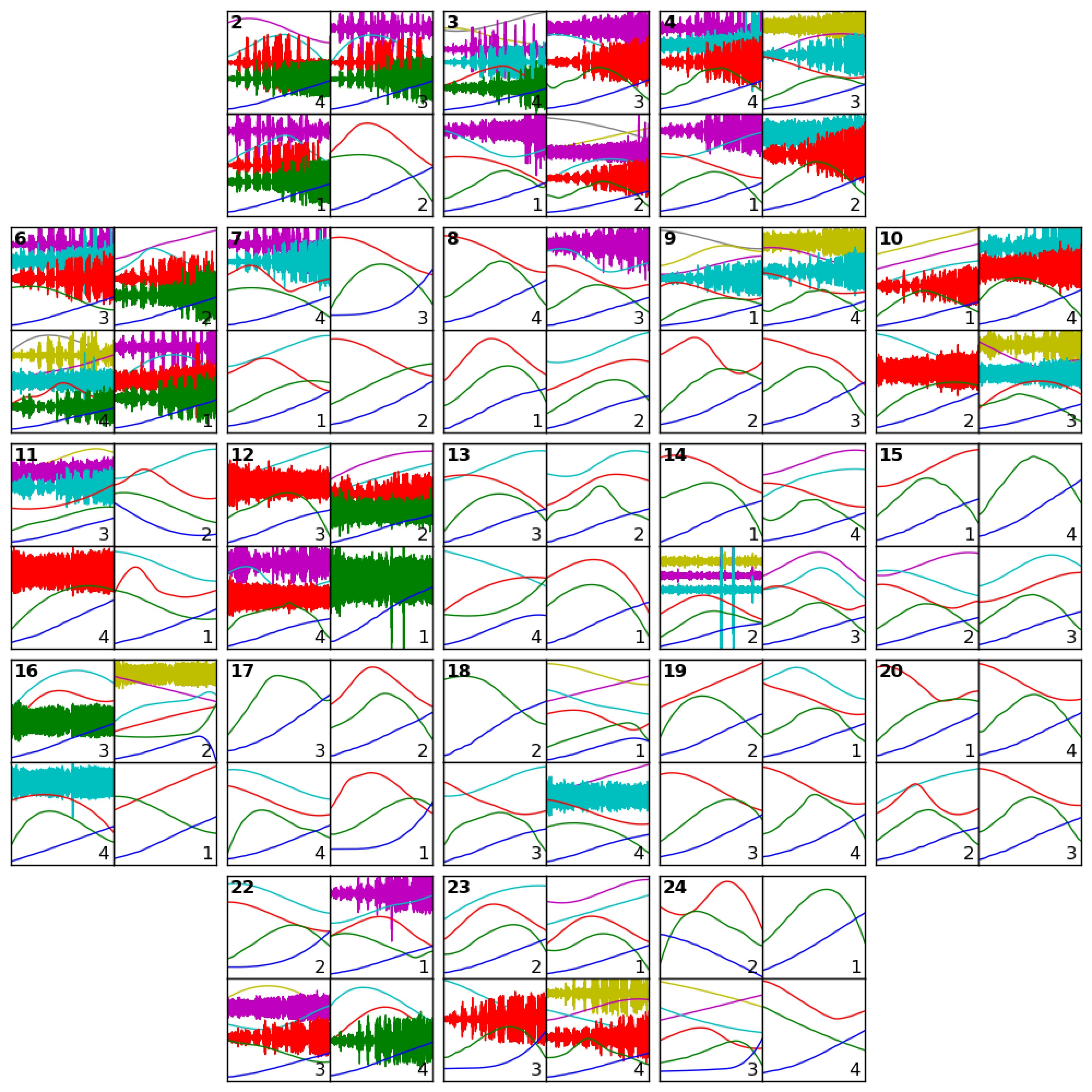}
  \caption{Q1 trend basis sets inferred for each module (labelled in
    bold at the top left of each group of plots) and output channel
    (labelled at the bottom right of each panel). The blue, green,
    red, cyan, magenta, yellow and grey lines show trends 1 to 7
    respectively (the number of trends identified varies). A vertical
    offset has been added between consecutive trends for clarity (the
    most important trend is always the lowest).}
  \label{fig:4x4trends}
\end{figure*}

We then repeated the trend identification step, first on a
module-by-module basis, then separately for each output channel. The
basis trends identified in the latter case are shown in
Figure~\ref{fig:4x4trends}. We compared the results from all three
approaches, and ultimately settled on the output channel specific
basis. The motivation for this is two fold. First, it is reasonable to
expect that the systematics affecting the data collected on different
detectors and read by different sets of electronics might differ, and
visual examination of the corrected light curves appeared to bear this
out, although the differences were very small.

Each of the Q1 light curves was then corrected using the set of basis
trends identified for the corresponding output channel. The full set
of basis trends and corrected light curves is available online at
www.physics.ox.ac.uk/users/aigrain/KeplerSys/.

\section{Discussion}
\label{sec:disc}

\begin{figure*}
  \includegraphics[width=0.49\linewidth]{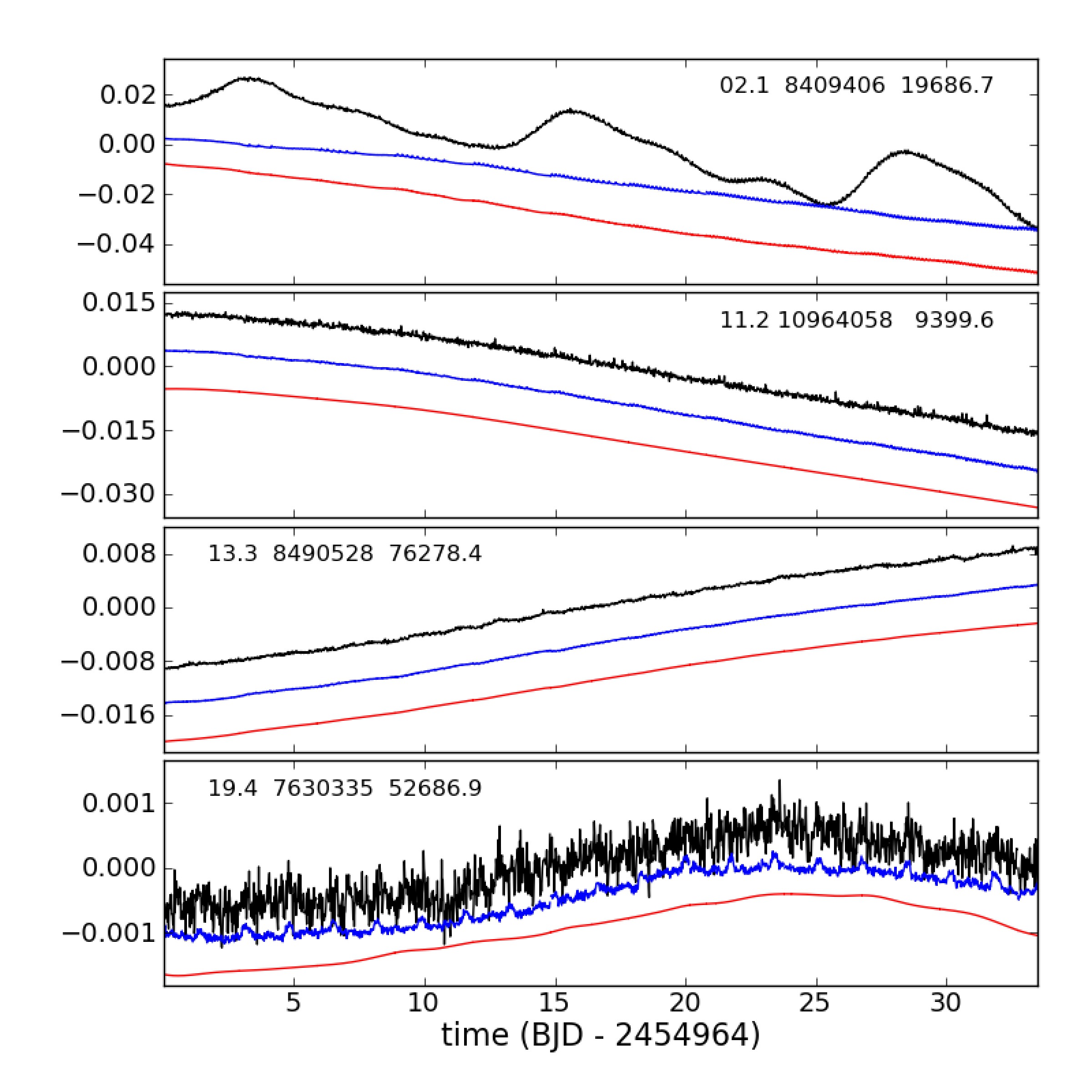} \hfill
  \includegraphics[width=0.49\linewidth]{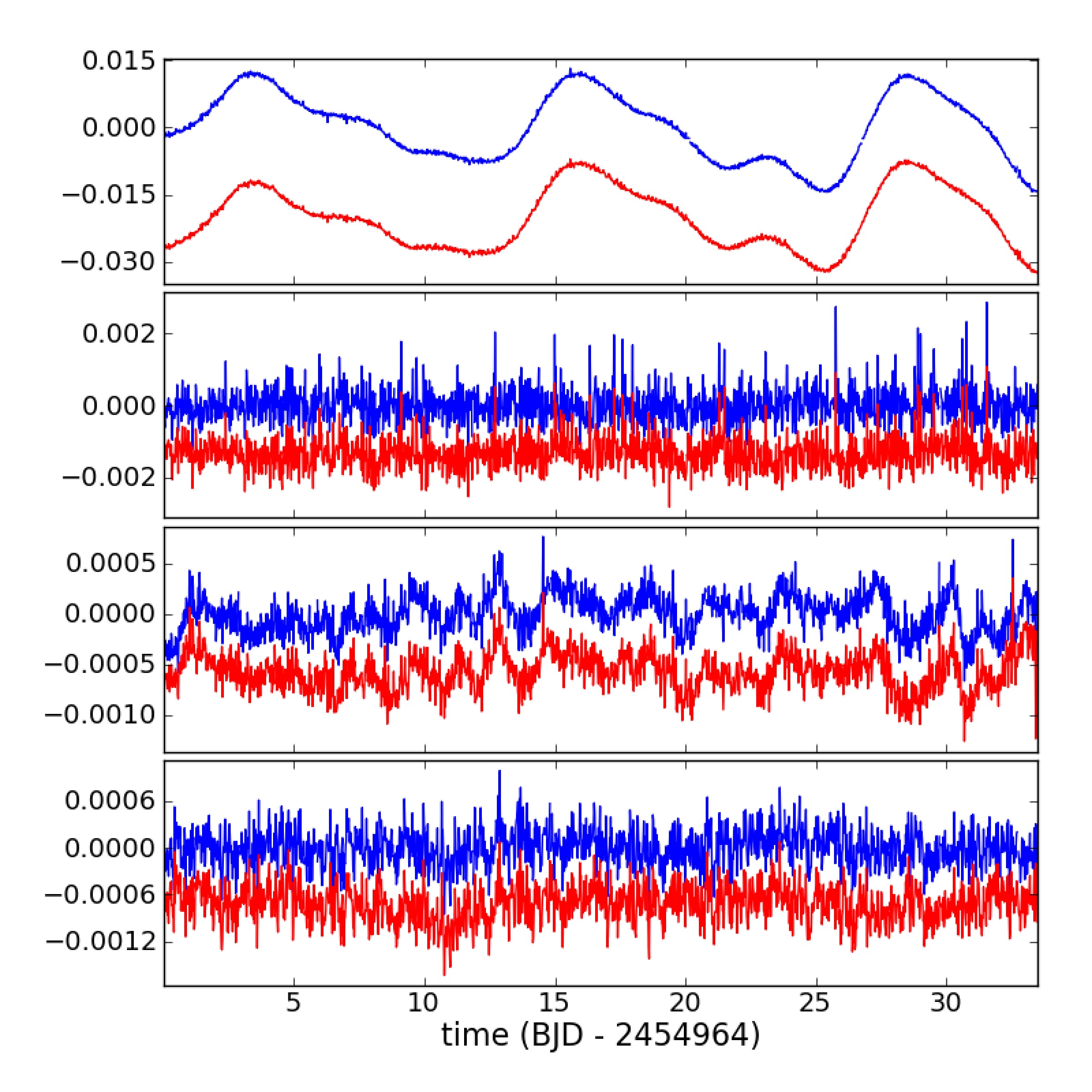}
  \caption{Example results from Q1. Each row corresponds to a
    particlar object. In the left column, the raw data and the
    corrections applied by the PDC-MAP and ARC are shown in black,
    blue and red, respectively. The module and output channel number,
    KIC number of the target, and median flux in the raw light curve,
    are also shown. In the right column, the PDC-MAP and ARC corrected
    data are shown in blue and red respectively. Each light curve has
    been normalised by dividing them by their median and subtracting
    unity. A vertical offset has also been applied between the curves
    in each panel, to improve clarity. }
  \label{fig:eg_results}
\end{figure*}
 
In this section, we assess the performance of the ARC systematics
correction on the Q1 data, and compare it to the PDC-MAP results.

Figure \ref{fig:eg_results} shows a selection of raw Q1 light curves
along with the PDC-MAP- and ARC-corrected versions. These were
selected at random on four different output channels in different
parts of the detector, and are intended to represent a fairly typical
subset of light curves, rather than a best- or worst-case
scenario. The most striking outcome of the comparison is that the
PDC-MAP and ARC results are, to first order, very similar. Both
successfully preserve stellar variability in most cases, and remove
most of the trends which one would deem to be systematic after
visually inspecting many light curves. 

\begin{figure*}
  \includegraphics[width=\linewidth]{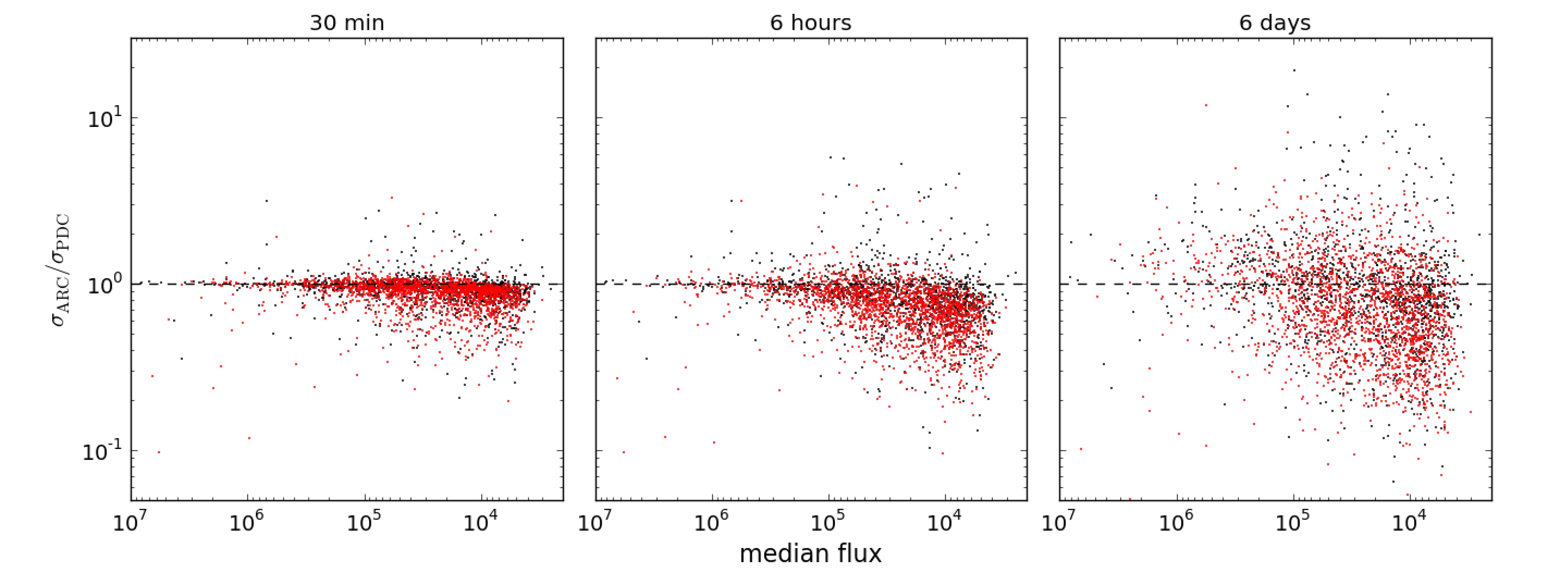} \hfill
  \caption{Comparison of the light curve scatter on different
    timescales after correction by PDC-MAP and by the ARC. Each panel
    shows the ratio of the light curve scatter resulting from the two
    corrections as a function of the median flux in the light
    curve. The x-axis has been reversed (so brighter stars are to the
    left) and plotted on a log scale (so it is similar to a magnitude
    scale). The black points correspond to output channel 1 on module
    2, where the `reaction wheel' trend was found to be proeminent,
    and the red points to output channel 1 on module 13, near the
    centre of the detector, where only long-term trends were
    identifed.}
  \label{fig:rms_comp}
\end{figure*}

However, there are some minor differences between the corrections
applied by the PDC-MAP and the ARC. In the top panel of
Figure~\ref{fig:eg_results}, the low-frequency component of the
applied corrections differs slightly. It is very difficult to
determine, on a dataset of this duration, which correction is more
appropriate. In the bottom panel, the PDC-MAP attempts to correct for
the effect of periodic pointing anomalies caused by the presence of an
eclipsing binary (EB) among the guide stars used in Q1. This is a
known effect, which was remedied after Q1 (by eliminating the
offending object from the set of guide stars). As we shall see below,
although the EB signature is present among the basis vectors
identified by the PDC-MAP pipeline, it remains challenging to
correct. The EB signature is not captured at all by the ARC. It is
visible in the candidate trends itendified for some (not all) of the
output channels, before the EMD denoising step, but it is never
represented in the intrinsic mode with the largest variance, which is
adopted as the `smoothed' basis trend. This is a limitation of our
approach. More importantly, however, in this example the PDC-MAP also
introduces significant amounts of high-frequency noise, which was not
present in the original light curve. The EMD de-noising step avoids
this, at the cost of failing to correct for the EB trend.
 
To investigate the differences between the PDC-MAP and ARC further, we
computed the relative scatter of the corrected light curves on a range
of timescales. We worked from the normalised light curves (after
dividing them by the raw median flux), smoothed them on the timescale
of interest using a median filter of the corresponding width, and
estimated the scatter $\sigma$ as 1.48 times the median of the
absolute deviations from unity\footnote{This diagnostic is equivalent
  to the standard deviation in the case of Gaussian distributed noise,
  but is less senstive to outliers
  \protect\citep{Hoa+83}.}. Figure~\ref{fig:rms_comp} shows the ratio
$\sigma_{\rm ARC}/\sigma_{\rm PDC}$ (where the subscript PDC refers to
the PDC-MAP) as a function of the median flux, for two representative
output channels (using all stars in each channel, namely 1290 and 2007
respectively), and for three different timescales: 30 minutes (the
sampling of Kepler long-cadence observations), 6 hours (most relevant
for transit detection, and 6 days (relevant, for example, to stellar
rotation studies).

On all timescales, the ratio is below 1 for the majority of stars,
meaning that the scatter of the ARC-corrected light curves is smaller
than that of the PDC-MAP light curves. This effect is particularly
noticeable for the fainter stars, where the scatters
resulting from the two corrections often differ by a factor of 3 or
more. At this stage, it would be tempting to interpret this as
indicating that the PDC-MAP introduces some spurious signal into many
of the light curves, which the ARC does not, presumably thanks to the
de-noising step. Of course, we cannot entirely rule out the
alternative possiblity that the ARC might be removing real
astrophysical signal. However, it seems implausible that it would do
systematically across a wide range of timescales, particularly because
the ARD priors used at the trend removal stage ensure that trends are
only removed if there is significant evidence for them in the data.

In order to verify that ARC is indeed removing only systematics,
and that any alterations of the underlying (astrophysical) signals are minimal,
we performed an injected signal test. We randomly selected 200 low-variability
stars, displaying little other than white noise in their ARC-processed light
curves, from mod2.out1. We added artificial sinusoidal signals (similar to those
described in Section 3.1) to the \emph{raw} light curves for these stars. We
then used the ARC to identify a basis from to a set of light curves including
some with injected signals and some without. This basis was then used to detrend
both sets of light curves, allowing us to recover an estimate of the injected
signals by subtraction. We then compared this to the actual injected signal in
each case: any discrepancies must have been introduced by the ARC process. To
quantify these discrepancies, 
we then evaluated, for each star, the variance of the residuals (the difference
between the estimated and injected signals), divided by the variance of the
injected signal. This is a quantitative measure of any discrepancy introduced by
the ARC. We found that this discrepancy measure has a mean value of 1\% and is
never above 5\%, confirming that the ARC procedure is unlikely to remove real
astrophysical variability.

We
also note that the number of cases, where the PDC-MAP and the ARC give
very different scatters, increases at longer timescales. There are a
few cases where the ARC scatter is significantly larger than the
PDC-MAP value, and these are more numerous in the module 2 example
than in the module 13 one.

\begin{figure}
  \includegraphics[width=\linewidth]{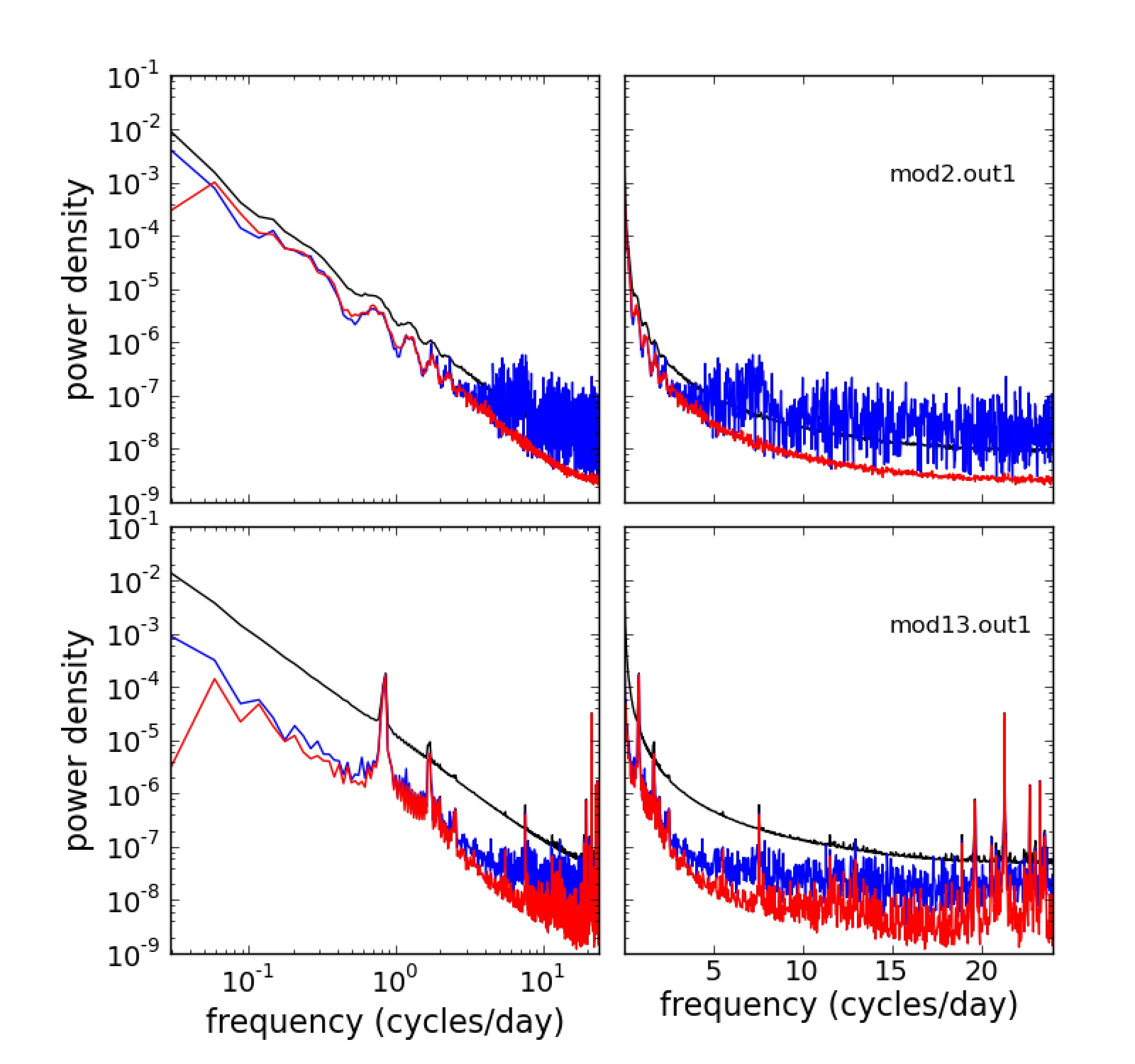} \hfill
  \caption{Comparison of the `average' power density spectra for
    bright stars before and after correction with the PDC-MAP and the
    ARC, in two representative output channels (top and bottom). The
    black, blue and red curves correspond to the raw, PDC-MAP and ARC
    data, respectively. The two columns show the same data, but the
    x-axis scale is logarithmic in one case, and linear in the other.}
  \label{fig:psd_comp}
\end{figure}
 
\citet{gilliland+:11} carried out an in-depth investigation of the
noise properties of the PDC data. In particular, they showed that the
majority of Sun-like stars observed by Kepler appear to display more
variability on 6.5-hour timescales than the Sun. This is an important
result, because it has a potentially serious impact on Kepler's
ability to detect transits of habitable planets. To estimate the
variability on transit timescale, \citet{gilliland+:11} used a
quantity known as the 6.5-hour combined differential photometric
precision (CDPP).  CDPP estimates computed by the Kepler pipeline
itself are now publicly available and distributed with the Kepler
data.  Our 6-hour $\sigma$ estimates are not exactly identical to the
CDPP values for the same dataset, because they are obtained in a
slightly different fashion, but they do measure a similar quantity,
namely the light curve scatter on typical transit timescales. Although
improving the detectability of transits was not our primary motive in
developing the ARC, the latter yields slightly lower scatter
on transit timescales than the PDC, and hence the ARC may prove useful
for transit searches.

We also computed `average' power spectra before and after correction,
for the bright stars (median raw flux $> 10^5$\,e$^{-}$/s) in the same
two output channels (giving 56 and 69 stars for the
  two output channels respectively). These are simply averages of the
individual power spectra of each light curve. The results are shown in
Figure~\ref{fig:psd_comp}. Again, both systematics correction methods
significantly reduce the light curve scatter on most timescales, and
their behaviour is very similar at low frequencies (except for the
very lowest, but these are poorly constrained due to the limited
duration of the dataset). However, the power spectra conclusively
demonstrate the fact that the PDC-MAP introduces high-frequency noise
into the light curves \citep[see also][though this reports results
using PDC rather than PDC-MAP]{murphy:12}. This effect is more
significant in some modules than in others, but the differences with
the ARC become noticeable upwards of about 2 cycles per
day. We note that, for mod2.out1, the set of
  systematics inferred by the ARC algorithm contains several
  high-frequency components. Similar trends may well be present in
  mod13.out1, but they are not significant in a large enough fraction
  of the light curves from that channel to be identified. This may
  explain the presence of excess power at high-frequencies in the
  average power spectrum of the ARC-corrected light curves from that
  channel, compared to mod2.out1. We ran a test where we arbitrarily
  fixed the number of trends identified to 5; this leads to the
  identification of the `reaction-wheel heater' trend in most channels, but
  with very low spectral radius, indicating that it is not significant
  in the majority of light curves. This highlights the capricious
  trade-off between over- and under-fitting: in this work we have
  attempted to find a balance using the spectral radius threshold, but
  other choices are possible.

The power spectra shown in the bottom panel of
Figure~\ref{fig:psd_comp} correspond to the central module of the
Kepler detector, where the only trends identified by the ARC were
long-term. There are clearly systematic effects at well-defined
frequencies, which are not well captured by either the PDC-MAP or the
ARC. The signal at $\sim 0.8$ cycles/day and harmonics thereof
correspond to the aforementioned EB guide star; there are also a
number of higher-frequency effects. The fact that neither the PDC-MAP
nor the ARC correct these effects well suggests that they may not obey
one of the fundamental assumptions underlying both methods. For
example, if the contribution of a given trend to a given light curve
varies during the quarter, the linear basis model will not be able to
capture it.

There are also some interesting differences in the average power at a
given frequency, and in the relative reduction in this power
introduced by the systematics corrections, for the two output channels
shown. However, it would probably be unwise to read too much into this,
as these power spectra are based on relatively small numbers of light
curves ($\sim 50$ in each case). Another point which we note as
interesting, but refrain from discussing further, is that all the
power spectra -- before and after systematics correction of any kind
-- are remarkably well-described by power laws with index $-2$, as
expected for an auto-regressive stochastic process, and frequently
used to model the so-called `solar background' (\citealt{har85}. See
\citealt{McQuillan+:12} for a more detailed discussion of this aspect).

\begin{figure*}
  \includegraphics[width=0.49\linewidth]{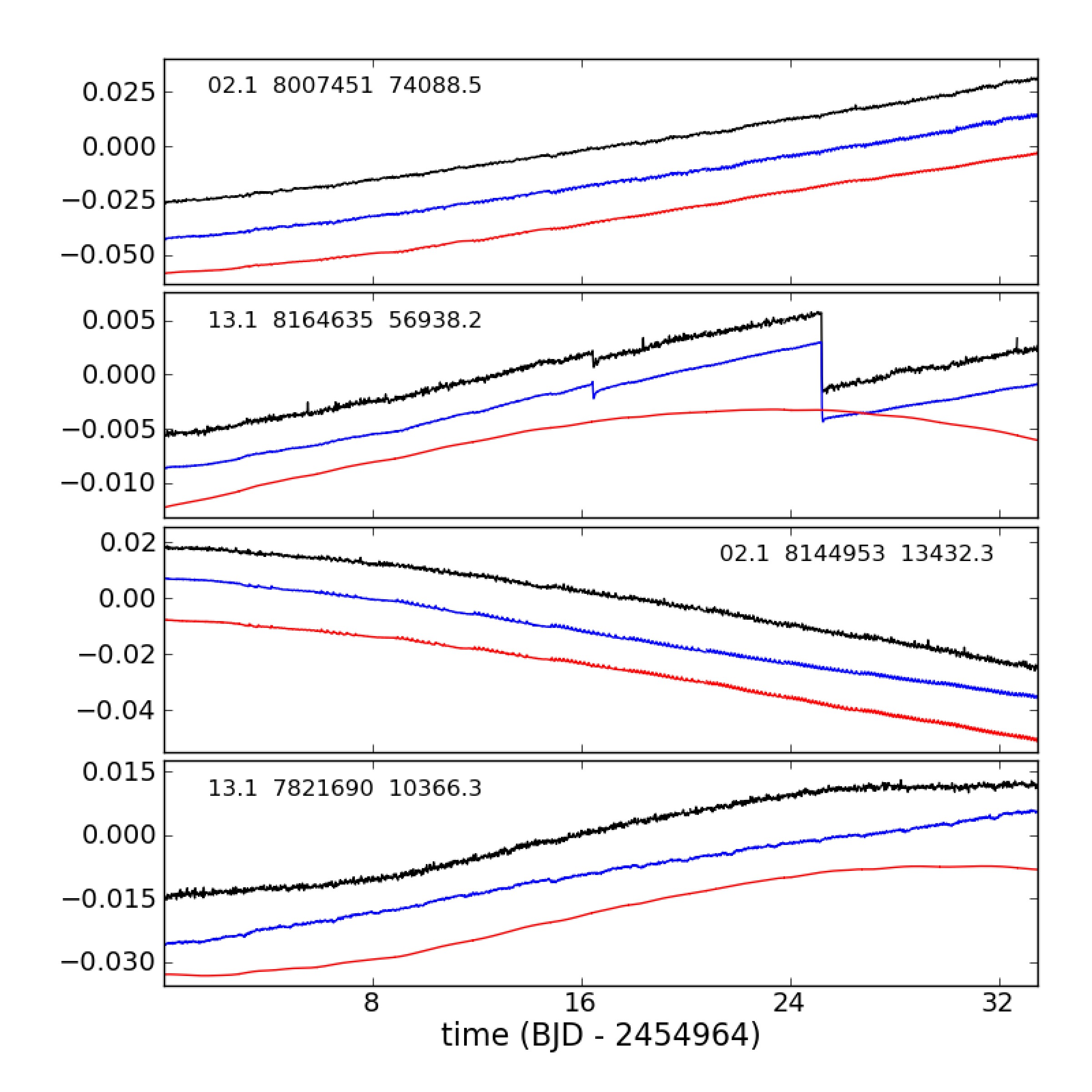} \hfill
  \includegraphics[width=0.49\linewidth]{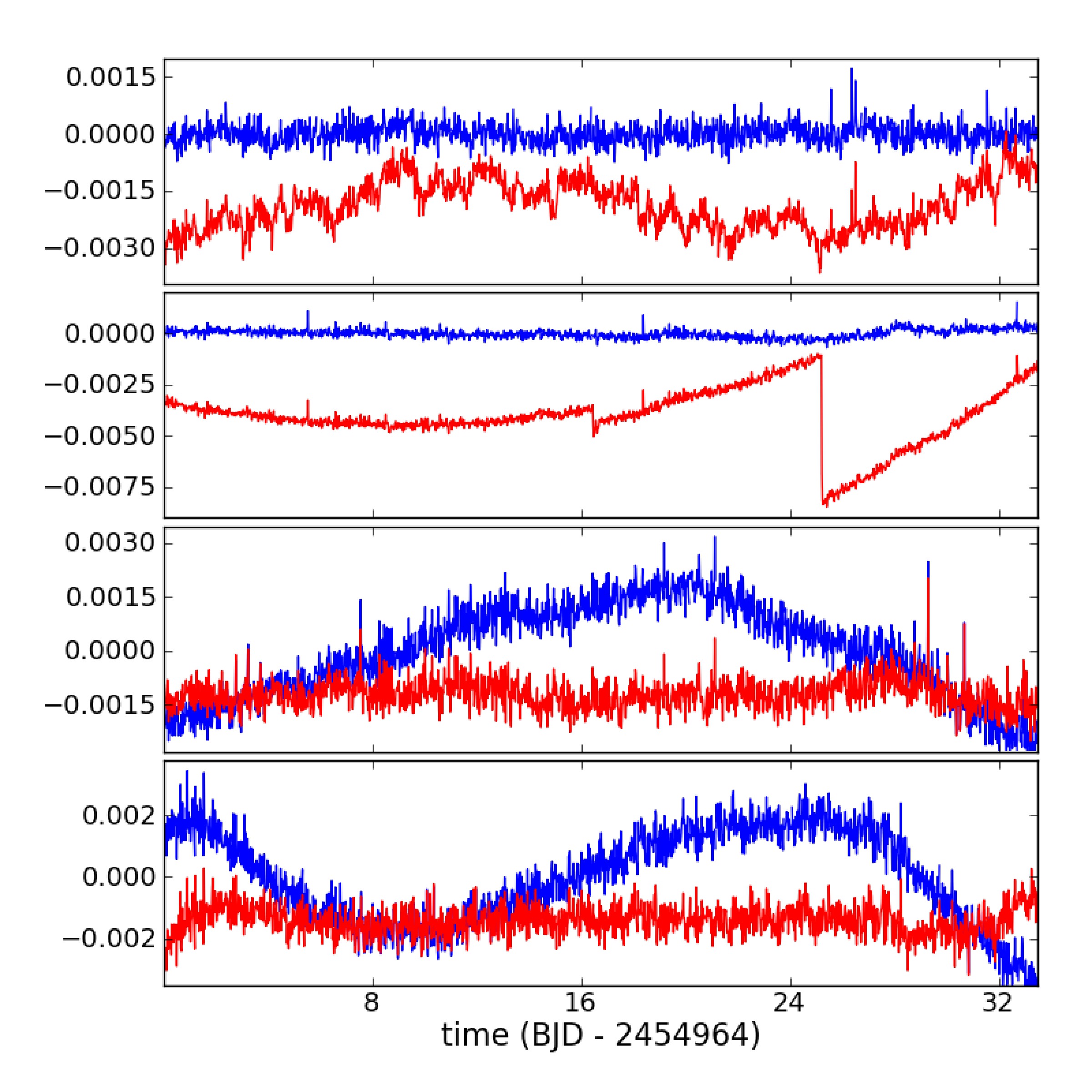}
  \caption{Example results from Q1 where there are significant deviations
between the ARC and PDC-MAP detrended light curves. Each row corresponds to a
    particlar object. In the left column, the raw data and the
    corrections applied by the PDC-MAP and ARC are shown in black,
    blue and red, respectively. The module and output channel number,
    KIC number of the target, and median flux in the raw light curve,
    are also shown. In the right column, the PDC-MAP and ARC corrected
    data are shown in blue and red respectively. Each light curve has
    been normalised by dividing them by their median and subtracting
    unity. A vertical offset has also been applied between the curves
    in each panel, to improve clarity. In the top two examples post-ARC scatter
is larger than that from PDC-MAP correction and in the lower two it is less.}
    \label{fig:arc_pdf_diff_eg}
  \end{figure*}

There are a few larger differences between the ARC and PDC-MAP
corrections. In Figure \ref{fig:arc_pdf_diff_eg} we highlight
representative, but rare, examples of large deviations. The top two
examples show cases where the post-ARC scatter is larger than the
post-PDC scatter (by a factor of $>2$ on 30 min timescales, $>3$ on 6
hr timescales and $>5$ on 6 day timescales). In both cases the light
curve contains a strong signature of an effect that the ARC cannot
deal with, for example the binary guide star (top row) or strong
discontinuities (pixel sensitivity drop-outs, 2nd row). The bottom two
examples show cases where the post-ARC scatter is considerably smaller
than the post-PDC scatter (by a factor of $>3$ on 30 min timescales,
$>5$ on 6 hr timescales and $>10$ on 6 day timescales). In both cases
this is because the range of basis vectors available in the PDC-MAP
detrending algorithm does not explain the light curve as effectively
as the ARC.

It is important to highlight two points regarding these deviations. Firstly, 
the cases where post-ARC scatter is significantly larger than the
post-PDC-MAP scatter are much rarer than \textit{vice versa}. Secondly,
the effects the ARC method cannot currently deal with are either not present in
later quarters (e.g. the eclipsing binary guide star) or will be dealt with
separately by utilising jump-correction software, developed primarily for
quarter 2 analysis, which will be described in a later paper.

\section{Conclusions}
\label{sec:conclusions}

We have presented a novel method to correct systematic trends present
in large ensembles of light curves, while preserving `true'
astrophyiscal variability, which we call the ARC method. The ARC is
specifically designed for continuous, high-precision observations,
where most stars display significant variability. It follows
established practice in using linear basis models, but incorporates a
number of new features, which are intended to ensure its robustness
(Bayesian framework, use of shrinkage priors), and its computational
efficiency (approximate inference using variational Bayes). A
de-noising step is used to prevent the introduction of spurious random
noise into the light curves.

We demonstrated the performance of the method on a simple synthetic
dataset, and on the data from the first month of Kepler
operations. The results were satisfactory in both
cases. In the synthetic example, the systematic trends
  were correctly identified. The original light curves included the
  building blocks of realistic stellar signals, namely harmonic functions with random, variable phases and amplitudes, and were  recovered robustly, with no evidence of systematic changes in
  amplitude or frequency. For the Kepler data, the ARC results are
similar to those of the PDC-MAP component of the Kepler pipeline for
most targets, but with some important differences. In particular, we
find that the PDC-MAP still tends to introduce
high-frequency noise into the light curves (just as
  noted by \citealt{murphy:12} for the original PDC). Importantly,
this effect remains noticeable on the timescales typical of planetary
transits. To address this problem, the Kepler team
  recently developed a new incarnation of the PDC, known as
  multi-scale MAP, or msMAP for short \citep{DR21}, which uses a
  band-splitting approach to separate systematics on different
  timescales. This does reduce the injection of high-frequency noise
  into the light curves compared to the PDC-MAP. However, in msMAP the
  treatment of long-term systematics (on timescales $\ge 21$ days) has
  reverted to the maximum likelihood approach of the original PDC
  pipeline. Consequently there is no protection against overfitting,
  and msMAP suppresses all variations on timescales longer than $\sim 20$ days --
  whether astrophysical or instrumental \citep[see Fig.~1 in][]{DR21}.
  Therefore, although we have not had the opportunity to make a
  detailed comparison between the new msMAP data and the ARC, we
  expect the letter to be preferrable for variability studies.

We did, however, identify certain types of systematics which are not
well-corrected by either the PDC-MAP or ARC. Further work needs to be
done on these: one area in particular which could be
  improved is the de-noising step, as the EMD method used for that
  step in the present paper is not suitable for sharp systematic
  effects. Another possibility, which we intend to investigate, is to
treat the systematics as multiplicative, rather than additive, by
working in log flux units. This which would account
  for gradual amplitude changes, which could occur due to the combined
  effect of pixel drift and aperture contamination (if one of the
  stars in the aperture is more variable than the other).  The impact
of the residual systematic effects should be relatively limited,
because they are confined to well-defined frequency ranges. Overall,
we therefore conclude that the ARC-corrected light curves could become
a very useful resource for both stellar astrophysics and exoplanet
searches, and we have made the Q1 data publicly
available\footnote{See www.physics.ox.ac.uk/users/aigrain/KeplerSys.}
(the code is also available on request).

However, the full potential of the Kepler data lies in its very long
baseline.  This paper focused on the first month of data only, whereas
more than three years are now in the public domain. A number of
additional difficulties arise when trying to process the additional
quarters, and to stitch together multiple quarters. Many of the
quarters contain discontinuities, most of which are associated with
monthly interruptions in the observations to transfer data back to
Earth. Many of the discontinuities are followed by quasi-systematic
trends which are present in most light curves, but with slightly
different shapes.  As the satellite rotates by $90^{\circ}$ about its
viewing axis between each quarter, each star falls on a different
detector, and this causes changes in both the median flux level and
the nature and amplitude of the systematic trends. The ARC is not
designed to address these problems. We are in the process of
developing a complementary set of `fault-correction' tools, to be
applied on a light-curve by-light-curve basis before running the ARC
on each channel. The same methodology can then be used to stitch
together data from different quarters. These developments will be
covered in a forthcoming paper. 

As a final cautionary note, we wish to stress that the ARC is
intended to provide a fast, automated and robust treatment of a
large ensemble of light curves. It is not intended to replace the
detailed examination of the pixel-level data \citep{PyKe,har+12}, which
remains the method of choice for the correction of systematic trends
and artefacts in small numbers of light curves. 

\section*{Acknowledgments}

We wish to thank the Kepler Science Operations Centre and pipeline teams, and in particular Jon Jenkins and Jeoffrey Van Cleve, for taking the time to discuss our methodology with us and providing some constructive feedback. AM and SA are supported by grants from the UK Science and Technology Facilities Council (refs ST/F006888/1 and ST/G002266/2). This paper includes data collected by the Kepler mission. Funding for the Kepler mission is provided by the NASA Science Mission Directorate. All of the data presented in this paper were obtained from the Mikulski Archive for Space Telescopes (MAST). STScI is operated by the Association of Universities for Research in Astronomy, Inc., under NASA contract NAS5-26555. Support for MAST for non-HST data is provided by the NASA Office of Space Science via grant NNX09AF08G and by other grants and contracts. SRo and SRe would like to thank the UK EPSRC for support under the ORCHID grant, EP/IO11587/1.

\appendix

\section{ Bayesian Linear Basis Model}
\label{sec:vblbm}
 
In a general linear basis model, the observations of a dependent
variable are modelled as a linear combination of basis functions of an
independent variable, plus a noise term. For consistency with the
notation used in the rest of the paper, we call the independent
variable $t$ and the dependent variable $d$, so that:
\begin{equation}
  d(t) =  \sum_{k=1}^{K} w_k \phi_k(t) + \epsilon,
  \label{eq:lbm}
\end{equation}
where the $w$'s are known as the \emph{factor weights} and the
$\phi$'s are the \emph{basis functions}. The noise term, $\epsilon$,
is taken to be drawn from a normal (Gaussian) distribution with zero
mean and precision (inverse variance) $\beta$. Note that an offset, or
\emph{bias} term, can easily be included in the model by including a
basis function which is one everywhere.

Defining the column vectors $\bm{w} \equiv [w_1$, \ldots, $w_K]\tp$ and
$\bmg{\phi}(t) \equiv [ \phi_1(t)$, \ldots, $\phi_K(t) ]\tp$,
we can re-write Equation~(\ref{eq:lbm}) as:
\begin{equation}
  d(t) =  \bm{w}\tp \bmg{\phi}(t) + \epsilon.
  \label{eq:lbmvec}
\end{equation}
We then obtain a system of simultaneous linear equations describing
the dataset as a whole:
\begin{equation}
  \bm{d} =  \bm{w}\tp \bmg{\Phi} + \bmg{\epsilon},
  \label{eq:lbmmat}
\end{equation}
where $\bm{\Phi}$ is an $N \times K$-element matrix with elements
$\phi_k(t_n)$ and $\bmg{\epsilon} \equiv [\epsilon_1$, \ldots,
$\epsilon_N]\tp$.

The \emph{likelihood}, i.e.\ the probability that our model gave rise
to the observations, is then simply 
\begin{equation}
p(d|\bm{w},\beta) = \Gauss{\bm{d};\bm{w}\tp \bmg{\Phi},\beta^{-1}
  \bm{I}}
\label{eq:lik}
\end{equation}
where $\Gauss{\bm{a}; \bm{b}, \bm{C}}$ is the probability of drawing a
vector $\bm{a}$ from a multi-variate Gaussian distribution with mean
vector $\bm{b}$ and covariance matrix $\bm{C}$, and $\bm{I}$ is the
identity matrix.

\subsection{Maximum Likelihood}
\label{sec:ML}

The maximum likelihood (ML) solution for the weights is given by the
standard pseudo-inverse equation, namely
\begin{equation}
\bm{w}_{\rm ML}  =  (\bmg{\Phi}^{\sf T}\bmg{\Phi})^{-1} \bmg{\Phi}^{\sf T} \bm{d}.
\label{eq:ML}
\end{equation}
However, the ML approach is prone to over-fitting.  Some relief from
the inherent problems associated with ML solutions may be obtained by
introducing a prior over the weights and obtaining the \emph{maximum a
  posteriori} (MAP) solution. This approach still relies on
\emph{point estimates} of the parameters of the models, and
consequently fails to take into account some of the intrinsic
uncertainties.

A full Bayesian solution is obtained by marginalising over the
posterior distributions of the variables, which can be achieved, for
example, using sample-based approaches such as Markov-Chain
Monte-Carlo (MCMC). These can be computationally intensive however,
and they scale poorly to large numbers of parameters, and it is often
problematic to establish whether convergence has been reached.  In
this paper, we advocate an alternative solution based on the
\emph{variational Bayes} (VB) framework.  In recent years VB has been
extensively used in the machine learning literature as a method of
choice for approximate Bayesian inference, as it offers computational
tractability even on very large data sets. A full tutorial on VB is
given in \cite{bishop,fox}. In what follows we describe the key
features and concentrate on the derivations of update equations for
the linear basis model at the core of this paper.

\subsection{The full Bayesian model}
\label{sec:vb_bayes_full}
In a Bayesian framework, the priors over the parameters of the model must be specified. For each of the weights, a zero-mean Gaussian prior with precision $\alpha$ is adopted. The zero mean ensures that 
the weight associated with a given basis function is only non-zero if the data requires it to be. The Gaussian is the least informative choice of distribution for a quantity that can be either positive or negative, and is therefore appropriate when little information is known about the weights a priori. It is also convenient to chose a form for the prior that is conjugate with the likelihood, because it makes the marginalization process analytic. The joint prior over all the weights is thus:
\begin{equation}
p(\bm{w} | \alpha) = \Gauss{\bm{w}; \bm{0}, \alpha^{-1} \bm{I}}.
\label{global}
\end{equation}
The value of $\alpha$ does not enter into the likelihood, but it does
control the parameters, and it is therefore known as a
hyper-parameter.  In a fully Bayesian treatment, rather than fixing
$\alpha$ to a specific value, we marginalise over it, under a
so-called hyper-prior. It is convenient to select a hyper-prior for
$\alpha$ that is conjugate with the prior over the weights,
because then the marginalisation can be done analytically. As the
prior is a Gaussian, and $\alpha$ should always be positive (because
it represents an inverse variance), a suitable form for the
hyper-prior is a Gamma distribution with shape parameter $a$ and scale
parameter $b$:
\begin{equation}
p(\alpha)  =  \mathcal{G}(\alpha; a, b) \equiv \frac{1}{\Gamma(a) b^a}
\, \alpha^{a-1} \, e^{-\alpha/b},
\end{equation}
where $\Gamma(x)$ is the Gamma function. 

The noise precision, $\beta$, is also a parameter of the model, for
which we need to adopt a prior. Again, a Gamma distribution is a
suitable choice, with hyperparameters\footnote{We discuss the choices
  for all our hyperparameters later in the Appendix.} $c,d$:
\begin{equation}
p(\beta)  =  \mathcal{G}(\beta; c, d).
\end{equation}

We concatenate all the parameters and hyperparameters of the model
into the vector $\bmg{\theta}=[\bm{w},\alpha,\beta]\tp$.  As the
weights depend upon the scale $\alpha$ but not the noise precision
$\beta$ the joint prior distribution over $\bmg{\theta}$ factorises as:
\begin{equation}
p(\bmg{\theta}) = p(\bm{w}|\alpha) \, p(\alpha) \, p(\beta).
\label{prior}
\end{equation}

Our objective in Bayesian regression is to estimate the posterior
distribution, which fully
describes our knowledge regarding the parameters of the model,
$\bmg{\theta}$, given the data $\bm{d}$. This is related to the
likelihood and prior by Bayes's theorem:
\begin{equation}
  p(\bmg{\theta}|\bm{d})
  = \frac{p(\bm{d} | \bmg{\theta}) \, p(\bmg{\theta})}{p(\bm{d})}.
\end{equation}
The denominator is the \emph{evidence} or \emph{marginal likelihood} of
the data under the model, and is given by
\begin{equation}
\label{eq:marg}
p(\bm{d}) = \int p(\bm{d}|\bmg{\theta}) \, p(\bmg{\theta})
\, \mathrm{d}\bmg{\theta} 
\end{equation}
More than a mere normalisation term, the evidence is a quantitative
measure of the extent to which the data supports the model. 

\section{Variational Bayesian Inference}
\label{sec:VB}

Whether we are interested in the model evidence or in the posterior
distribution over individual parameters (after marginalising over the
others), the functional dependence of the likelihood and posterior
distribution on the parameters are generally unknown. However, we can
propose an analytical approximation for any one of them, which can if
necessary be refined later. This gives rise to the class of
\emph{approximate inference} methods. 

In \emph{variational} inference, we introduce an analytically tractable
distribution $q(\bmg{\theta}|\bm{d})$, which we use to approximate the posterior
distribution $p(\bmg{\theta}|\bm{d})$. We use this to write the log
evidence as the sum of two separate terms:
\begin{eqnarray}
  \log p(\bm{d})  & = & \log p(\bm{d}) \int q(\bmg{\theta}|\bm{d}) \,
  \mathrm{d}\bmg{\theta} \\
  & = & \int q(\bmg{\theta}|\bm{d}) \, \log p(\bm{d}) \,
  \mathrm{d}\bmg{\theta} \\
  & = & \int q(\bmg{\theta}|\bm{d}) \, \log \left[
    \frac{p(\bm{d},\bmg{\theta})}{p(\bmg{\theta}|\bm{d})}
  \right] \, \mathrm{d}\bmg{\theta} \\
  & = & \int q(\bmg{\theta}|\bm{d}) \, \log \left[
    \frac{p(\bm{d},\bmg{\theta})}{p(\bmg{\theta}|\bm{d})}
    \cdot \frac{q(\bmg{\theta}|\bm{d})}{q(\bmg{\theta}|\bm{d})}
  \right] \, \mathrm{d}\bmg{\theta} \\
  & = & \int q(\bmg{\theta}|\bm{d}) \log
\frac{p(\bm{d},\bmg{\theta})}{q(\bmg{\theta}|\bm{d})}
~\mathrm{d}\bmg{\theta} \nonumber \\ 
  & & \hspace{1cm} - \int q(\bmg{\theta}|\bm{d}) \log
\frac{p(\bmg{\theta}|\bm{d})}{q(\bmg{\theta}|\bm{d})}
~\mathrm{d}\bmg{\theta} \\
\label{eq:VB}
 & \equiv & \mathcal{F}(p,q) + {\rm KL}(p,q).
\label{fundamental}
\end{eqnarray}
Equation \ref{fundamental} is the fundamental equation of the
VB-framework. The first term on the right-hand side,
$\mathcal{F}(p,q)$, is known as the (negative) variational free
energy. The second, ${\rm KL}(p,q)$ is the Kullback-Leibler (KL)
divergence between the approximate posterior $q$ and the true
posterior $p$.  Importantly, the KL divergence is always
positive. $\mathcal{F}(p,q)$ thus provides a strict \emph{lower bound}
on the log evidence. Moreover, because the KL divergence is zero when
the two densities are the same, $\mathcal{F}(p,q)$ will become equal
to the log evidence when the approximating posterior is equal to the
true posterior, i.e.\ when
$q(\bmg{\theta}|\bm{d})=p(\bmg{\theta}|\bm{d})$.
The aim of VB learning is therefore to maximise $\mathcal{F}(p,q)$ and
so make the approximate posterior as close as possible to the true
posterior. This requires the extremisation of an integral with respect
to a functional, which is typically achieved using the \emph{calculus
  of variations}. 

However, to obtain a \emph{practical} inference algorithm, we restrict
the range of proposal posterior distributions over which we perform
the optimization. First, we restrict ourselves to distributions
belonging to the exponential family. This ensures that extremisation
over the \emph{function} $q(\bmg{\theta})$ can be replaced exactly by
extremisation with respect to the \emph{parameters} $\bmg{\theta}$. As
this family includes all the commonly-used probability distributions,
this constraint is not normally problematic.  What we obtain is then a
set of coupled update equations over the parameters which are cycled
until a convergence criterion is met. This approach is a
generalisation of the \emph{expectation-maximisation} (EM) algorithm,
which is obtained as a special case of variational Bayes in the limit
of the $q()$ distributions being replaced by delta functions at their
maximum-likelihood values \citep{bishop}.

Furthermore, we also assume that the posterior distribution is \emph{separable},
meaning that it can be written as a product of independent functions
of different parameters (or subsets of the parameters). This makes the resultant inference algorithm computationally very rapid, with little loss of information.

We begin by writing this factorisation as:
\begin{equation}
  q(\bmg{\theta}|\bm{d}) = \prod_i q(\theta_i |\bm{d}).
 \label{eq:separable}
\end{equation}
This makes the resultant inference algorithm computationally very rapid, with little loss of information. The aim is to find the distribution $q(\bmg{\theta}|\bm{d})$ for which $\mathcal{F}(p,q)$ is largest, which is done by optimising each of the factors in turn. We start by substituting this separable expression into
$\mathcal{F}(p,q)$:
\begin{equation}
\label{eq:Fqp1}
\mathcal{F}(p,q) = \int \prod_i q(\theta_i |\bm{d}) \left[
\log p(\bm{d},\bmg{\theta}) - \sum_i \log q(\theta_i |\bm{d}) 
\right] \mathrm{d}\bmg{\theta}.
\end{equation}
Next, we write down separately the terms which depend on one of the parameters,
$\theta_j$:
\begin{eqnarray}
\mathcal{F}(p,q) &=&
\iint \mathrm{d} \theta_j \mathrm{d}\bmg{\theta_{(i \neq j)}} \, q(\theta_j |\bm{d}) \prod_{i \neq j} q(\theta_i |\bm{d}) \ \\
&\times& \left[ \log p(\bm{d},\bmg{\theta}) - \log q(\theta_j |\bm{d}) -
  \sum_{i \neq j} \log q(\theta_i |\bm{d}) \right]. \nonumber
\end{eqnarray}
We then rearrange the above equation, so as to isolate the dependency on $\theta_j$:
\begin{eqnarray}
  \mathcal{F}(p,q) &=&\int q(\theta_j |\bm{d}) \, 
  \left\{ \int \log p(\bm{d},\bmg{\theta}) \, \right. \nonumber \\
  &\times& \left. \prod_{i \neq j}  q(\theta_i |\bm{d}) \, \mathrm{d}\bmg{\theta_{(i \neq j)}} \right\}
  \, \mathrm{d} \theta_j
 \nonumber \\
 &  - &  \int q(\theta_j |\bm{d}) \, \log q(\theta_j |\bm{d}) \, \mathrm{d} \theta_j \,
   \int \prod_{i \neq j} q(\theta_i |\bm{d}) \,
  \mathrm{d}\bmg{\theta_{(i \neq j)}} \nonumber \\
  &-&   \int \prod_{i \neq j} q(\theta_i |\bm{d}) 
  \sum_{i \neq j} \log q(\theta_i |\bm{d}) \,
  \mathrm{d}\bmg{\theta_{(i \neq j)}} \nonumber \\
 &\times& \int q(\theta_j |\bm{d}) \, \mathrm{d} \theta_j. 
\label{eq:3parts}
\end{eqnarray}
The second factor in each of the last two terms in Equation~(\ref{eq:3parts}) is
the integral of a probability distribution, and is thus equal to
one. Furthermore, the third term is independent of $\theta_j$, so:
\begin{eqnarray}
  \mathcal{F}(p,q) & = & \int q(\theta_j |\bm{d}) \, 
  \left\{ \int \log p(\bm{d},\bmg{\theta}) \, \prod_{i \neq j}
    q(\theta_i |\bm{d}) \, \mathrm{d}\bmg{\theta_{(i \neq j)}} \right\}
  \, \mathrm{d} \theta_j
 \nonumber \\
  & - &
  \int q(\theta_j |\bm{d}) \, \log q(\theta_j |\bm{d}) \, \mathrm{d}
  \theta_j \, + {\mbox{const}}(j)
\label{eq:2parts}
\end{eqnarray}
where ${\mbox{const}}(j)$ represents a term that is constant with
respect to $\theta_j$.  The quantity in curly braces in the first term
of Equation~(\ref{eq:2parts}) is the expectation of $\log
p(\bm{d},\bmg{\theta})$ under the candidate posterior distribution for
all the other parameters, $\prod_{i \neq j} q(\theta_i |\bm{d})$. We
wish to isolate the part of it which depends on $\theta_j$. To this
end, we define the quantity
\begin{equation}
f(\theta_j) \defn \int \log \left[ p(\bm{d},\bmg{\theta}) \right]_j \, \prod_{i \neq j}
    q(\theta_i |\bm{d}) \, \mathrm{d}\bmg{\theta_{(i \neq j)}},
\end{equation}
where $\left[ p(\bm{d},\bmg{\theta}) \right]_j$ contains all the terms
in $p(\bm{d},\bmg{\theta})$, which depend on $\theta_j$. 
We can then write
\begin{equation}
\int \log p(\bm{d},\bmg{\theta}) \, \prod_{i \neq j}
    q(\theta_i |\bm{d}) \, \mathrm{d}\bmg{\theta_{(i \neq j)}} 
= f(\theta_j) + {\mbox{const}} (j)
\end{equation}
and thus Equation~(\ref{eq:2parts}) becomes:
\begin{eqnarray}
  \mathcal{F}(p,q) & = & \int q(\theta_j |\bm{d}) \, \left[ f(\theta_j) + {\mbox{const}} (j)
  \right] \, \mathrm{d} \theta_j \nonumber\\
 & - &  \int q(\theta_j |\bm{d}) \, \log q(\theta_j |\bm{d}) \, \mathrm{d}
  \theta_j \, + {\mbox{const}}(j).
\label{eq:1part}
\end{eqnarray}
The non-constant terms in Equation \ref{eq:1part} can readily be
identified as the negative KL divergence between $q(\theta_j |\bm{d})$
and $\exp \left[ f(\theta_j) + {\mbox{const}}(j) \right]$. This divergence is minimised (and is zero)
when the two distributions are the same. Therefore, the free energy
$\mathcal{F}(p,q)$ is maximised with respect to $\theta_j$ simply by
setting 
\begin{equation}
\label{eq:up1}
\log q^{\rm{new}}(\theta_j |\bm{d}) = f^{\rm{old}}(\theta_j) + {\mbox{
  const}}(j).
\end{equation}
The additive constant in Equation \ref{eq:up1} is set by normalising
the distribution $q^{\rm{new}} (\theta_j |\bm{d})$. Thus, if we take the
exponential of both sides and normalize, we obtain the updated
distribution for parameter $\theta_j$ as:
\begin{equation}
\label{eq:up2}
  q^{\rm{new}}(\theta_j|\bmg{d}) = \frac{\exp \left[f^{\rm{old}}(\theta_j)\right
    ]}{\int \exp \left [ f^{\rm{old}}(\theta_j) \right ] \mathrm{d}\theta_j}.
\end{equation}
As previously mentioned, we have chosen to use candidate distributions which belong to the exponential
family, so that extremisation over the function $q(\bmg{\theta}|\bm{d})$ can
be replaced exactly by extremisation with respect to the parameters
$\bmg{\theta}$, and thus we can re-write Equation \ref{eq:up2}
\begin{equation}
\label{eq:up3}
  q(\theta_j^{\rm{new}}|\bmg{d}) = \frac{\exp \left[f(\theta_j^{\rm{old}})\right
    ]}{\int \exp \left [ f(\theta_j^{\rm{old}}) \right ] \mathrm{d}\theta_j}.
\end{equation} 
We note that, as with the EM algorithm, each iteration is
\emph{guaranteed} to improve the marginal likelihood of the data
under the model.

In the following subsections, we describe how the parameters of the
linear basis model (the weights $\bm{w}$, the weight precision
$\alpha$, and the noise precision $\beta$) are inferred using variational Bayes and detail the update
equations associated with each of them. 

\subsection{Updating the weights $\bm{w}$}

The marginal for the weights is
\begin{equation}
f(\bm{w})  =  \iint q(\beta |\bm{d}) \, q(\alpha|\bm{d}) \log
\left[p(\bm{d}|\bm{w},\beta) \,
p(\bm{w}|\alpha) \right] \, \mathrm{d}\alpha \, \mathrm{d}\beta 
\label{eq:f_w}
\end{equation}
Our update equations are hence of the form 
\begin{equation}
q(\bm{w}|\bm{d}) \propto \exp[f(\bm{w})]
\end{equation}
Substituting for the terms in Equation \ref{eq:f_w} gives
\begin{eqnarray}
  f(\bm{w}) & = & - \int q(\beta|\bm{d}) \, \frac{\beta}{2} \,
  (\bm{d}-\bmg{\Phi}\bm{w})\tp (\bm{d}-\bmg{\Phi}\bm{w}) \, \mathrm{d}\beta \nonumber \\
  & & - \int q(\alpha|\bm{d}) \, \frac{\alpha}{2} \, \bm{w}\tp \bm{w} \, \mathrm{d}\alpha \nonumber \\
  & = & - \frac{\hat{\beta}}{2} \, (\bm{d}-\bmg{\Phi}\bm{w})\tp
  (\bm{d} - \bmg{\Phi}\bm{w}) -\frac{\hat{\alpha}}{2} \, \bm{w}\tp\bm{w}
\end{eqnarray}
where $\hat{\alpha}$ and $\hat{\beta}$ are the expectations of the
weight and noise precisions under the proposal distributions over
$\alpha$ and $\beta$ (see the next two sections for details of the form of these distributions). The weight posterior is therefore a normal distribution: $
q(\bm{w}|\bm{d})= \Gauss{\bm{w};\hat{\bm{w}},\hat{\bm{\Sigma}}}$, with
\begin{eqnarray}
\hat{\bm{w}} & = & \hat{\bm{\Sigma}} \, \hat{\beta} \, \bmg{\Phi}\tp \bm{d} \nonumber \\
\hat{\bm{\Sigma}} & = & (\hat{\beta} \, \bmg{\Phi}\tp \bmg{\Phi}  + \hat{\alpha} \,\bm{I})^{-1}
\label{eq:wUpdates}
\end{eqnarray}
Thus, the posterior precision matrix,
$\hat{\bm{\Sigma}}^{-1}$, takes the usual Bayesian form of being the
sum of the data precision, and the prior precision,
$\hat{\alpha}\bm{I}$. If $\hat{\alpha}=0$, i.e. in the absence of
prior on the weights, we recover the ML solution of Equation~(\ref{eq:ML}).

\subsection{Updating the weight precision, $\alpha$}

We let
\begin{eqnarray}
f(\alpha) &=& \iint q(\beta | \bm{d}) \, q(\bm{w}|\bm{d}) \log
\left[p(\bm{w}|\alpha) \, p(\alpha) \right] \, \mathrm{d}\bm{w} \, \mathrm{d}\beta \nonumber \\
& = & \int q(\bm{w}|\bm{d}) \log \left[p(\bm{w}|\alpha)\, p(\alpha) \right] \mathrm{d}\bm{w}
\label{eq:wPrec}
\end{eqnarray}
As before, the negative free energy is maximised when
\begin{equation}
q(\alpha|\bm{d}) \propto \exp[f(\alpha)]
\end{equation}
By substituting for the terms in Equation \ref{eq:wPrec} we find that
the updated weight precision posterior density is a Gamma distribution
$q(\alpha|\bm{d})= \mathcal{G}(\alpha;\hat{a},\hat{b})$ where the
updated hyper-hyperparameters, $\hat{a}$ and $\hat{b}$, are given by
\begin{eqnarray}
  \frac{1}{\hat{a}} & = &  \hat{\bm{w}}\tp \hat{\bm{w}} + \frac{1}{2} \, {\rm
    Tr}(\hat{\bm{\Sigma}}) + \frac{1}{a_0} \\ \nonumber
\hat{b} & = & \frac{K}{2} + b_0, \\
\label{eq:alphaUpdates1}
\end{eqnarray}
where $a_0$ and $b_0$ are the initial estimates for $a$ and $b$, and
$K$ is the number of basis functions.  The updated
value for $\alpha$, to be substituted into
Equations~(\ref{eq:wUpdates}), is then simply the mean of this
distribution:
\begin{equation}
\hat{\alpha}  =  \hat{a} \hat{b}.
\label{eq:alphaUpdate2}
\end{equation}

\subsection{Updating the noise precision, $\beta$}

Again we start by writing the marginal
\begin{eqnarray}
  f(\beta) & = & \iint q(\alpha|\bm{d}) \, q(\bm{w}|\bm{d}) \log
  [p(\bm{d}|\bm{w},\beta) \, p(\beta)] \, \mathrm{d}\bm{w} \, \mathrm{d}\alpha \nonumber \\
  &=& \int q(\bm{w}|\bm{d}) \log [p(\bm{d}|\bm{w},\beta) \, p(\beta)] \,\mathrm{d}\bm{w}.
\label{eq:noisePrec}
\end{eqnarray}
The negative free energy is then maximised when
\begin{equation}
q(\beta|\bm{D}) \propto \exp[f(\beta)]
\end{equation}
By substituting for the terms in Equation \ref{eq:noisePrec} we
find, as with $\alpha$, that the posterior distribution over $\beta$
is of Gamma form, $q(\beta|\bm{d}) =
\mathcal{G}(\beta;\hat{c},\hat{d})$. The update equations for
$\hat{c}$, $\hat{d}$ and $\hat{\beta}$ are:
\begin{eqnarray}
  1/\hat{c} &=& \frac{1}{2} \, (\bm{d} - \bmg{\Phi} \hat{\bm{w}})\tp
  (\bm{d}-\bmg{\Phi}\hat{\bm{w}}) + \frac{1}{2} \, {\rm Tr} \left (
    \hat{\bmg{\Sigma}} \,\bmg{\Phi}\tp \bmg{\Phi} \right ) + \frac{1}{c_0} \nonumber \\
\hat{d} &=& \frac{N}{2} + d_0 \nonumber \\
\hat{\beta} &=& \hat{c} \, \hat{d}
\label{eq:betaUpdates}
\end{eqnarray}
where $N$ is the number of data points.

\subsection{Structured priors}

Instead of using the isotropic Gaussian of Equation \ref{global},
where the distribution over all the weights has a common scale
(defined by the single hyperparameter $\alpha$), we can split the
weights into groups and allow different groups to have different
scales in their distributions; each weight $w_i$ can indeed have its
own scale hyperparameter.  This approach is often referred to as
\emph{Automatic Relevance Determination} (ARD)
\citep{Neal98assessingrelevance,bishop}, because by inspecting the
inferred scales associated with the weights we can see which (groups
of) weights are relevant to the problem at hand. The posteriors for
the weights of any basis functions which are not helpful in explaining
the data will evolve towards zero-mean distributions with vanishingly
small variance. Conversely, the weights of basis functions which are
well supported by the data will entertain larger variances in their
posterior distributions.  This means we may operate with a rich basis
set and allow the Bayesian model to select only those basis functions
that have explanatory power in the data.

We can allow different weights to have different scales, but still
take into account domain knowledge which -- for example -- may lead us
to believe that certain parameters should have a similar posterior
scale, by adopting \emph{structured priors}
\citep{Penny_Roberts_MAR:02}, of the form
\begin{equation}
p(\bm{w} | \{\alpha_g\}) = \prod_{g=1}^G \left(\frac{\alpha_g}{2\pi}\right)^{K_g/2} \exp (-\alpha_g
\, E_g(\bm{w}))
\label{ard}
\end{equation}
where the weights have been split into $G$ groups, with $K_g$
weights in the $g^{\rm th}$ group, 
\begin{equation}
E_g(\bm{w}) =  \frac{1}{2}\bm{w}\tp \bm{I}_g \bm{w} 
\end{equation}
and $\bm{I}_g$ is a diagonal matrix with ones in the row
corresponding to the $g^{\rm th}$ group, and zeros elsewhere.
Use of structured priors results in VB updates for the posterior
weight covariance and weight precision as follows
\begin{eqnarray}
\hat{\bm{\Sigma}} & = & (\hat{\beta} \, \bmg{\Phi}\tp \bmg{\Phi}  + \sum_{g=1}^G \hat{\alpha}_g \bm{I}_g)^{-1} \nonumber \\
1/\hat{a}_g & = &  E_g(\hat{\bm{w}}) +
\frac{1}{2} \, {\rm Tr}(\bm{I}_g \, \hat{\bm{\Sigma}} \, \bm{I}_g) + \frac{1}{a_0} \nonumber \\
\hat{b}_g & = & \frac{K_g}{2} + b_0 \nonumber \\
\hat{\alpha}_g & = & \hat{a}_g \hat{b}_g.
\end{eqnarray}
The other updates are exactly the same as for the global variance scale over the parameters.

\subsection{Implementation details}
All the code for the ARC was written in Matlab. The basis discovery phase for
all mod.out took just under an hour running on a quad-core 2.6GHz processor
under a Unix operating system. The subsequent trend removal over all mod.out
took some five minutes and scales linearly with the number of light curves being
detrended. Without loss of generality we normalise each light curve to have zero
mean and unit variance prior to analysis. We note that this transformation is
reversible subsequent to processing. The variational Bayes basis model, detailed
in this Appendix, is used in both the trend discovery and trend removal phases
of the algorithm, differing only in the details of the structured priors, as
described below.

\paragraph*{Trend discovery:} In this phase of the algorithm we implement
Equation \ref{eq:linMod} in which successive light curves are modelled as linear
combinations of one another. In this phase we use a single global weight
precision hyperparameter, $\alpha$, over which we place a Gamma distribution
with
hyper-hyperparameters $a_0=10^{-2}, b_0=10^{-4}$. The noise precision, $\beta$
is also Gamma distributed with initial hyper-hyperparameters
$c_0=10^{-2},d_0=10^{-4}$.

\paragraph*{Trend removal:} In this phase we wish to allow individual precisions
associated with the weights of the model, as each weight is linked to a putative
(pre-discovered) trend component. This allows shrinkage on trends that are not
present in a light curve and avoids falsely removing any components for which
there is not compelling evidence. We hence follow the procedure for structured
priors as detailed in the previous section of the Appendix. Once more we set
vague priors for each weight precision, $\alpha_g$, and for the noise precision
$\beta$, choosing $a_0=10^{-2},b_0=10^{-4},c_0=10^{-2},d_0=10^{-4}$.

\bsp

\label{lastpage}


\begin{thebibliography}{99}

\bibitem[\protect\citeauthoryear{Baglin et al.}{2009}]{CoRoT} 
Baglin A., Auvergne M., Barge P., Deleuil M., Michel E., CoRoT Exoplanet 
Science Team, 2009, IAUS, 253, 71 

\bibitem[\protect\citeauthoryear{Bakos et al.}{2002}]{Bakos+:02} Bakos
  G.~{\'A}., et al., 2003, PASP, 114, 974

\bibitem[\protect\citeauthoryear{Bakos et al.}{2007}]{Bakos+:07} Bakos
  G.~{\'A}., et al., 2007, ApJ, 670, 826 

\bibitem[\protect\citeauthoryear{Bishop}{2006}]{bishop} Bishop C.~M.,
  2006, \emph{Pattern Recognition and Machine Learning}, Springer

\bibitem[\protect\citeauthoryear{Borucki et al.}{2010}]{Borucki+:10}
  Borucki W.~J., et al., 2010, Sci, 327, 977
   
\bibitem[\protect\citeauthoryear{Everson \& Roberts}{2000}]{Everson+Roberts:00}
  Everson R, Roberts S, 2000, IEEE Trans. Sig. Proc. 48(7), 2083.

\bibitem[\protect\citeauthoryear{Fox \& Roberts}{2011}]{fox}
  Fox C, Roberts S, 2011, Artificial Intelligence Review, 38(2), 85

\bibitem[\protect\citeauthoryear{Gilliland et
    al.}{2011}]{gilliland+:11} Gilliland R.~L., et al., 2011, ApJS,
  197, 6

\bibitem[\protect\citeauthoryear{Harrison et al.}{2012}]{har+12}
  Harrison, T.~E., Coughlin, J.~L., Ule, N.~M., L{\' o}p{\`
    e}z-Morales, M., 2012, AJ, 143, 4

\bibitem[\protect\citeauthoryear{Harvey}{1985}]{har85} Harvey,
  J.~W. 1985, in \emph{Future missions in Solar, heliospheric and
    space plasma physics}, ESA SP-235

\bibitem[\protect\citeauthoryear{Hoaglin, Mostellar \&
    Tukey}{1983}]{Hoa+83} Hoaglin D.~C., Mostellar F., Tukey J.~W.,
  1983, \emph{Understanding Robust and Exploratory Data Analysis},
  John Wiley, New York

\bibitem[\protect\citeauthoryear{Huang et al.}{1998}]{Huang} 
 Huang N.~E., et al., 1998, RSPSA, 454, 903 

\bibitem[\protect\citeauthoryear{Jenkins et al.}{2010}]{Jenkins+:10}
  Jenkins J.~M., et al., 2010, ApJ, 713, L87 

\bibitem[\protect\citeauthoryear{Jenkins et al.}{2013}]{DataChar}
 Jenkins, J.~M., et al., 2013, Kepler data characteristics handbook,
 KSCI-19040-004

\bibitem[\protect\citeauthoryear{Kinemuchi et 
al.}{2012}]{PyKe} Kinemuchi K., Barclay T., Fanelli M., 
Pepper J., Still M., Howell S.~B., 2012, PASP, 124, 963 

\bibitem[\protect\citeauthoryear{Kov{\'a}cs, Bakos, \&
    Noyes}{2005}]{Kovacs+:05} Kov{\'a}cs G., Bakos G., Noyes R.~W.,
  2005, MNRAS, 356, 557

\bibitem[\protect\citeauthoryear{McQuillan, Aigrain, \&
   Roberts}{2012}]{McQuillan+:12} McQuillan A., Aigrain S., Roberts
  S., 2012, A\&A, 539, A137

\bibitem[\protect\citeauthoryear{Murphy}{2012}]{murphy:12} Murphy 
S.~J., 2012, MNRAS, 422, 665 

\bibitem[\protect\citeauthoryear{Neal}{1998}]{Neal98assessingrelevance}
  Neal R., 1998, in \emph{Neural Networks and Machine Learning},
Springer-Verlag, 97

\bibitem[\protect\citeauthoryear{Penny \& Roberts}{2002}]{Penny_Roberts_MAR:02}
 Penny W., Roberts S., 2002, in \emph{IEE Proceedings on Vision, image and signal processing},
   149(1), 33
   
\bibitem[\protect\citeauthoryear{Petigura \& Marcy}{2012}]{Petigura+Marcy:12}
Petigura E. and Marcy G., 2012, PASP 124, 1073.

\bibitem[\protect\citeauthoryear{Pollacco et al.}{2006}]{Pollacco+:06}
  Pollacco D.~L., et al., 2006, PASP, 118, 1407

\bibitem[\protect\citeauthoryear{Shannon}{1951}]{shannon_ent} Shannon
  C.~E., 1951, The Bell System Technical Journal, 30, 50

\bibitem[\protect\citeauthoryear{Smith et al.}{2012}]{PDC-MAP1} 
Smith J.~C., et al., 2012, PASP, 124, 1000 

\bibitem[\protect\citeauthoryear{Stumpe et al.}{2012}]{PDC-MAP2} 
Stumpe M.~C., et al., 2012, PASP, 124, 985 

\bibitem[\protect\citeauthoryear{Tamuz, Mazeh, \&
    Zucker}{2005}]{Tamuz+:05} Tamuz O., Mazeh T., Zucker S., 2005,
  MNRAS, 356, 1466 
  
\bibitem[\protect\citeauthoryear{}{2013}]{DR21} Thompson S.~E.,
J.~L. Christiansen, J. Jenkins, M.~R. Haas, 2013, Kepler data release
notes 21, KSCI-19061-00l

\bibitem[\protect\citeauthoryear{Udalski et al.}{2002}]{Udalski+:02}
  Udalski A., et al., 2002, AcA, 52, 1

\bibitem[\protect\citeauthoryear{Van Cleve \&
    Caldwell}{2012}]{kepInstHandbook} Van Cleve J.~E. \& Caldwell
  D.~A, 2009, \emph{Kepler Instrument Handbook} (KSCI-19033).

\bibitem[\protect\citeauthoryear{Wall et al.}{2003}]{PracStatAst} Wall
  J.~V., Jenkins C.~R., Ellis R., Huchra J., Kahn S., Rieke G.,
  Stetson P.~B., 2003, \emph{Practical Statistics for Astronomers},
  Cambridge University Press


\end{thebibliography}
\end{document}